\documentclass[showpacs,aps,prb,notitlepage,twocolumn,superscriptaddress]{revtex4-1}

\usepackage[utf8]{luainputenc}
\usepackage{lmodern}
\usepackage[svgnames, table]{xcolor}
\usepackage{graphicx}
\usepackage{enumerate}
\usepackage{tabulary}
\usepackage{dcolumn,bbding}
\usepackage{tabularx,hhline,array}
\usepackage{array}
\usepackage{multirow}
\usepackage{lipsum}
\usepackage{subfigure}
\usepackage{amsmath, amsfonts, amssymb}
\usepackage{wasysym,mathrsfs,euscript,eufrak}
\usepackage{tikz,graphicx}
\usepackage{pdfsync}
\usepackage{natbib}
\usepackage{hyperref}
\hypersetup{
colorlinks = True,
linkcolor = red,
urlcolor = blue,
citecolor = blue
}
\newcolumntype{C}{>{$}c<{$}}

\begin{document}

\title{Generic nearest-neighbour kagome model:\\ XYZ \& Dzyaloshinskii-Moriya, and comparison to pyrochlore}
\author{Karim Essafi}
\email{karim.essafi@oist.jp}
\affiliation{Okinawa Institute of Science and Technology Graduate University, Onna-son, Okinawa 904-0495, Japan}
\author{Owen Benton}
\email{john.benton@riken.jp}
\affiliation{Okinawa Institute of Science and Technology Graduate University, Onna-son, Okinawa 904-0495, Japan}
\affiliation{RIKEN Center for Emergent Matter Science (CEMS), Wako, Saitama, 351-0198, Japan}
\author{L. D. C. Jaubert}
\email{ludovic.jaubert@u-bordeaux.fr}
\affiliation{Okinawa Institute of Science and Technology Graduate University, Onna-son, Okinawa 904-0495, Japan}
\affiliation{CNRS, Univ. Bordeaux, LOMA, UMR 5798, F-33400 Talence, France}

\begin{abstract}
The kagome lattice is a paragon of geometrical frustration, long-studied for its
association with novel ground-states including spin liquids.
Many recently synthesized kagome materials feature rare-earth ions, which may be
expected to exhibit highly anisotropic exchange interactions.
The consequences of this combination of strong exchange anisotropy and
extreme geometrical frustration are yet to be fully understood.
Here, we establish a general picture of the interactions and resulting ground-states arising from nearest neighbour exchange anisotropy on the kagome lattice.
We determine a generic anisotropic exchange Hamiltonian from symmetry arguments. In the high-symmetry case where reflection in the kagome plane is a symmetry of the system, the generic nearest-neighbour Hamiltonian can be locally defined as a XYZ model with out-of-plane Dzyaloshinskii-Moriya interactions. We proceed to study its phase diagram in the classical limit, making use of an exact reformulation of the Hamiltonian in terms of irreducible representations (irreps) of the lattice symmetry group.
This reformulation in terms of irreps naturally explains the three-fold mapping between spin liquids recently studied on kagome by the present authors [\href{https://www.nature.com/articles/ncomms10297}{Nature Communications \textbf{7}, 10297 (2016)}].
In addition, a number of unusual states are stabilised in the regions where different forms of ground-state order compete, including a stripy phase with a local $\mathbb{Z}_8$ symmetry and a classical analogue of  a chiral spin liquid.
This generic Hamiltonian also turns out to be a fruitful hunting ground for coexistence of different forms of magnetic order, or of order and disorder, which we find is a particular property of the kagome lattice arising from the odd number of spins per frustrated unit.
These results are compared and contrasted with those obtained on the pyrochlore lattice, and connection is made with recent progress in understanding quantum models with $S=1/2$.
\end{abstract}

\date{\today}

\maketitle

%======================================================
%======================================================
%======================================================
\section{Introduction}

When confronted with a new magnetic material, one of the early questions is often to search for its microscopic Hamiltonian. Magnetic interactions are governed by a set of rules. In particular, they have to respect the symmetry of the lattice. For example, Dzyaloshinskii-Moriya (DM) interaction is a well-known consequence of the absence of an inversion centre between pairs of spins~\cite{Dzyaloshinsky1958,Moriya1960}. Hence, for any material, an analysis of its lattice symmetry provides a useful tool in determining a microscopic model~\cite{Curnoe2007}. Such symmetry-based approach has proven remarkably successful for a systematic parameterisation and understanding of rare-earth pyrochlore materials~\cite{Ross2011,Savary12b,Chang12a,Hayre13a,Oitmaa13a,Guitteny13b,Robert15a,Jaubert15b,Yan17a}, as well as for the Ba$_{3}$Yb$_{2}$Zn$_{5}$O$_{11}$ breathing pyrochlore~\cite{Rau16a,Haku16a,Savary16a} and YbMgGaO$_{4}$ triangular spin-liquid candidate~\cite{Li15a,Li16b,Shen16a,Paddison17a}. These successes are due, to some extent, to the nature of the rare-earth ions. Their $4f$ valence electrons give rise to short-range superexchange which can often be modeled by nearest-neighbour couplings, and thus require a limited number of coupling parameters. Additionally, their strong spin-orbit coupling facilitates anisotropic interactions~\cite{Rau15a}, providing the microscopic ingredients for exotic magnetic orders and textures.

In kagome materials, while the traditional Heisenberg antiferromagnet, for both classical~\cite{Chalker1992,Zhitomirsky2008,Chern13b} and quantum~\cite{Hermele2008,Yan2011,Iqbal2011,Lauchli16a,He16a,Liao17a} spins, has been investigated in great depth, the focus has lately shifted towards more anisotropic models. The experimental motivation does not only stem from rare-earth (R) compounds -- \textit{e.g.} R$_{3}$Ru$_{4}$Al$_{12}$ ternary intermetallic~\cite{Gorbunov14a,Ge14a,Nakamura15a,Chandragiri16a} or R$_{3}$Mg$_{2}$Sb$_{3}$O$_{14}$ tripod kagome~\cite{Dun16a,Scheie16a,Paddison16b,Dun17a} -- but also from copper-~\cite{Rigol2007,Zorko08a,Rousochatzakis09a,Shawish10a,Han12a,Zorko13a,Rousochatzakis2015} and iron-~\cite{Chernyshev15a,Chernyshev15b} based materials. On the theoretical front, anisotropy also offers a natural setting for spin-liquid ground states~\cite{Dodds13a,Schaffer13a,Lauchli15a,He15a,Carrasquilla15a,Essafi16a,Messio16a}.

The goal of this paper is to explore the zero-temperature phase diagram of the generic nearest-neighbour Hamiltonian allowed by the symmetry of the kagome lattice for classical Heisenberg spins.

We shall first explain in detail how to derive this Hamiltonian [Eqs.~(\ref{eq:J01gen}) - (\ref{eq:J20gen}) and (\ref{eq:J01}) - (\ref{eq:J20})], from the 
point group symmetry; see section \ref{sec:ham} and especially section \ref{sec:XYZDMham} for a non-technical summary and section \ref{sec:related} for comparison of our Hamiltonian with related generic models. 
The kagome symmetry allows for six independent coupling parameters. 
This can be reduced to four in the presence of a mirror symmetry in the plane of the kagome lattice itself.
In this latter case we have a Hamiltonian with four parameters
$\{J_{x},J_{y},J_{z},D\}$ which are best rationalised as an XYZ model with Dzyaloshinskii-Moriya, denoted XYZDM.
The XYZDM Hamiltonian is then diagonalized in section \ref{sec:irrep}, making use of the decomposition into irreducible representations (irreps). The irrep decomposition provides the natural order parameters for $\mathbf{q}=0$ long-range order on kagome \cite{Okuma17a}, as expressed in Eqs.~(\ref{eq:Order-Parameter-A1}-\ref{eq:Order-Parameter-Ec}) and represented in Fig.~\ref{fig:irrep}. In the basis formed by these order parameters, the XYZDM Hamiltonian becomes quadratic [section \ref{sec:diagham}]. This rewriting is exact and not a mean field approximation, which allows for a simple and exact determination of the ground-states for most of the phase diagram, as explained in section \ref{sec:howGS}. Sections \ref{sec:ham} and \ref{sec:irrep} closely follow the procedure developed for pyrochlore lattices in Refs.~[\onlinecite{Curnoe2007,McClarty2009,Benton2014,Yan17a}].
%
%======================
\begin{figure}[t]
\centering\includegraphics[width=5cm]{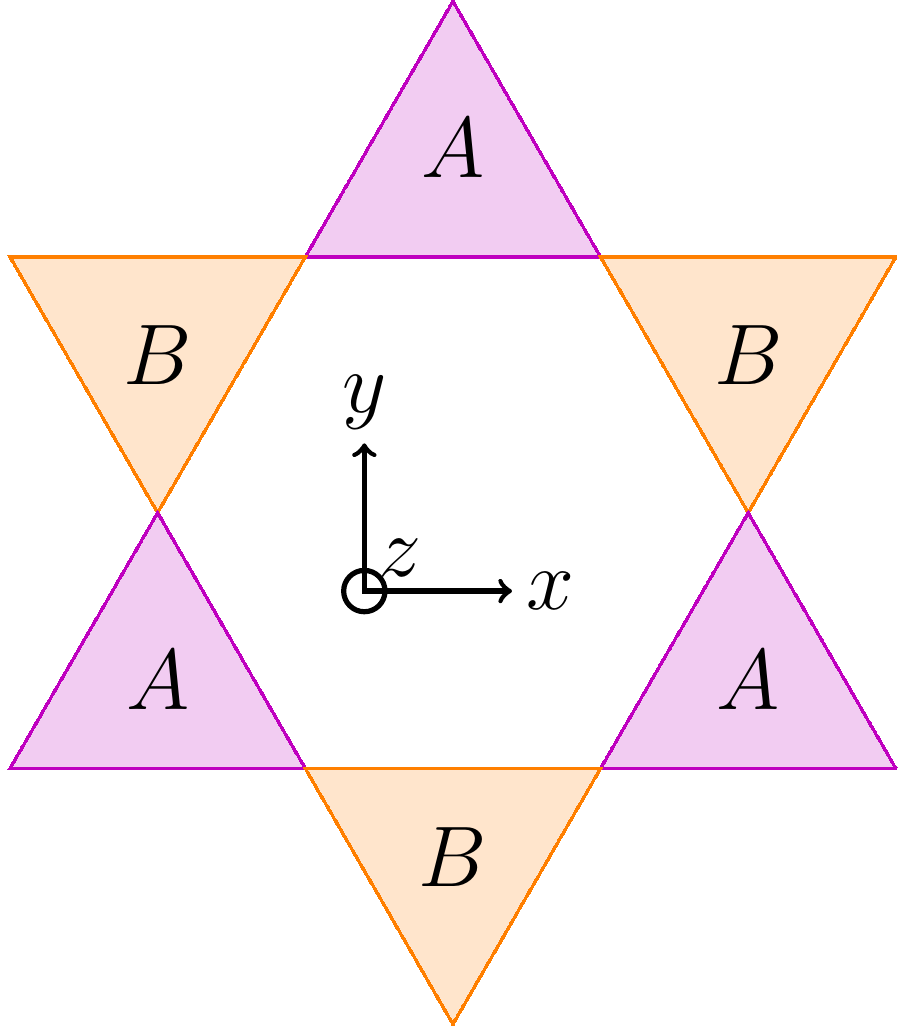}
\caption{The up and down triangles of the kagome lattice are respectively colored in violet (A) and orange (B). The $x$ and $y$ axes are in plane, while the $z$-axis is out of plane, pointing towards the reader.}
\label{fig:kagAB}
\end{figure}
%======================

A key difference between kagome and pyrochlore though is that some of the order parameters derived from the irrep decomposition on kagome correspond to non-physical states, in the sense that they describe configurations in which spins which are not of unit length. In practice, this means that some regions of the phase diagram in parameters space $\{J_{x},J_{y},J_{z},D\}$ support ground-states where different kinds of orders co-exist, or that a partial order of the spin degrees of freedom may co-exist with magnetic disorder. This complexity largely disappears for a portion of the XYZDM model where O(2) invariance is imposed in the kagome plane; this is the XXZ model with Dzyaloshinskii-Moriya \cite{Chernyshev15b}, noted XXZDM. Section \ref{sec:XXZDM} is devoted to the XXZDM model whose zero-temperature phase diagram is given in Fig.~\ref{fig:PhD}. Using the irrep decomposition, particular attention will be paid to the network of classical spin liquids supported by this model \cite{Essafi16a}; how it emerges from the surrounding ordered phases~\cite{Miyashita86a,Kuroda95a,Elhajal2002} and possibly connects to quantum spin liquids \cite{Lauchli15a,He15a,He2015,Zhu15a,Hu15a,Lauchli16a}. In section \ref{sec:XYZDM}, the condition of O(2) invariance is lifted and one recovers the XYZDM model. After discussing the inherent invariance and chiral asymmetry of the XYZDM model [sections \ref{sec:invA1A2b} and \ref{sec:chiasym}], we will describe a variety of specific Hamiltonians with ordered and disordered ground-states that, to the best of our knowledge, have not been observed before. In particular, an extended region of the XYZDM phase diagram supports ground-states with (i) local $\mathbb{Z}_{8}$ degeneracy and global stripe order [section \ref{sec:Z8}] and (ii) classical chiral spin liquids that can be mapped onto different tricolor problems [section \ref{sec:tricolor}]. In section \ref{sec:Mach}, we explicitly compare the analogies and differences between the generic models on kagome and pyrochlore lattices, before concluding in section \ref{sec:discuss}.

%======================================================
%======================================================
%======================================================
\section{Derivation of the generic nearest-neighbour kagome model}
\label{sec:ham}

%======================================================
%======================================================
\subsection{Which interactions are allowed on kagome ?}
\label{sec:J?}
The kagome lattice being made of corner-sharing triangles, any nearest-neighbour Hamiltonian can be written as a sum over triangles $X$
\begin{align}
\mathcal{H} = \sum\limits_{X \in A} \mathcal{H}_{\Delta}^A[X] + \sum\limits_{X \in B} \mathcal{H}_{\Delta}^B[X] 
\end{align}
where $A$ and $B$ refer to the sets of up and down triangles respectively [Fig.~\ref{fig:kagAB}].

\begin{figure}[b]
\includegraphics[width=\columnwidth]{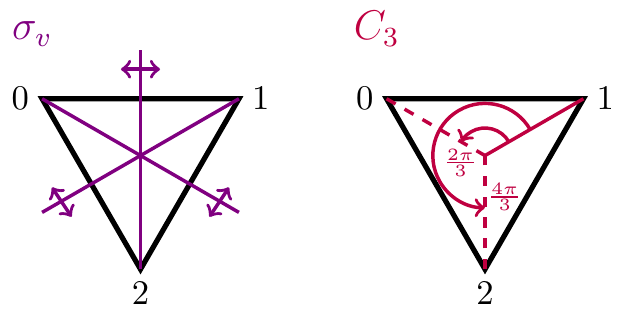}
\caption{
Graphical representations of the $\sigma_{v}$ reflections and $C_{3}$ rotations in the $C_{3v}$ symmetry group. The sublattices on a triangle, $\{S_0,S_1,S_2\}$, are labeled in the clockwise convention. Because of the antisymmetric Dzyaloshinskii-Moriya interaction [Eq.~(\ref{eq:JbasisB})], it is important to conserve the same clockwise convention for all triangles.
}
\label{fig:D3sym}
\end{figure}

Let $\hat{J}^{X}_{ij}$ be the coupling matrix between a pair of classical Heisenberg spins $\mathbf{S}_{i}$ and $\mathbf{S}_{j}$ on a $X\in\{A,B\}$ triangle, where the spins have unit length $|\mathbf{S}|=1$, and $i,j\in\{0,1,2\}$ label the kagome sublattices as defined in Fig.~\ref{fig:D3sym}. $A$ and $B$ triangles are related by lattice inversion $\hat{\mathcal{I}}$
\begin{eqnarray}
\hat{J}^{B}_{ij}=\hat{\mathcal{I}} \; \hat{J}^{A}_{ij} \; \hat{\mathcal{I}}.
\label{eq:}
\end{eqnarray}
Using the facts that the spins are axial vectors and thus invariant under lattice inversion,
\begin{eqnarray}
\hat{\mathcal{I}} \;{\bf S}_i={\bf S}_i\;,
\label{eq:SinvS}
\end{eqnarray}
one obtains for any pair of spins
\begin{align}
{\bf S}_i \; \hat{J}^{A}_{ij} \; {\bf S}_j &= {\bf S}_i \; \hat{\mathcal{I}}^2  \hat{J}^{A}_{ij} \; \hat{\mathcal{I}}^2 \;{\bf S}_j \nonumber \\
&= {\bf S}_i \; \hat{\mathcal{I}} \; \hat{J}^{A}_{ij} \; \hat{\mathcal{I}} \; {\bf S}_j \nonumber \\
&={\bf S}_i \; \hat{J}^{B}_{ij} \; {\bf S}_j\nonumber\\
\Rightarrow \hat{J}^{A}_{ij} &= \hat{J}^{B}_{ij}.
\end{align}
%
%======================
%======================
%
Hence, the Hamiltonian $\mathcal{H}_{\Delta}$ is the same for A and B triangles
\begin{align}
\mathcal{H}_{\Delta}^A = \mathcal{H}_{\Delta}^B = \mathcal{H}_{\Delta} = \sum_{\langle ij\rangle}^{} \sum_{\alpha, \beta}^{} S_i^\alpha \hat{J}_{ij}^{\alpha \beta} S_j^\beta,
\label{eq:Hamiltonian-Triangle}
\end{align}
where the Greek and Latin indices respectively label the spin components and the sublattices of a triangle [see Figs.~\ref{fig:kagAB} and \ref{fig:D3sym} for the labelling convention]. Alternatively, $\mathcal{H}_{\Delta}$ can be written in the form of a $9 \times 9$ coupling matrix $\mathcal{\hat{J}}$
\begin{align}
\mathcal{H}_{\Delta}
= \frac{1}{2} \tilde{S} \begin{pmatrix}
0 & \hat{J}_{01} & \hat{J}_{02} \\
\hat{J}_{10} & 0 & \hat{J}_{12} \\
\hat{J}_{20} & \hat{J}_{21} & 0
\end{pmatrix} \tilde{S}
= \frac{1}{2} \tilde{S} \mathcal{\hat{J}} \tilde{S},
\label{eq:9x9}
\end{align}
where $\tilde{S}$ is a $9$-component vector containing the magnetic degrees of freedom for each triangle
\begin{align}
\tilde{S} = (S_0^x,S_0^y,S_0^z,S_1^x,S_1^y,S_1^z,S_2^x,S_2^y,S_2^z).
\label{eq:repres}
\end{align}
By definition, the coupling matrix $\mathcal{\hat{J}}$ equals its transpose, $\mathcal{\hat{J}}=\mathcal{\hat{J}}^{t}$. This leaves $9\times 3=27$ parameters \textit{a priori} undetermined in the coupling matrix $\mathcal{\hat{J}}$. However, since we are interested in models respecting the symmetry of the kagome lattice, the coupling matrix $\mathcal{\hat{J}}$ must be invariant under action of the reflection and rotation symmetries illustrated in Fig. \ref{fig:D3sym}.
This corresponds to the group $C_{3v}$ which is of order $6$ and its group elements are the neutral element 
$e$, two $2n\pi/3$-rotations around the out-of-plane axis, and three reflections through the planes normal to each bond [Fig.~\ref{fig:D3sym}]
\begin{align}
C_{3v} = \left\{e, C_3^{i=1,2}, \sigma_v^{i=0,1,2}\right\}.
\label{eq:D3}
\end{align}
We shall treat the spins as transforming like axial vectors.
All $C_{3v}$ elements can be obtained by successive actions of one of the $C_{3}$ rotations and one of the reflections.

Representing these operations as $9 \times 9$ matrices in the basis defined by Eq. (\ref{eq:repres}) it is sufficient to consider
for example
\begin{eqnarray}
{\Gamma}(\sigma_{v}^{2})=
\begin{pmatrix}
0&0&0&1&0&0&0&0&0\\
0&0&0&0&-1&0&0&0&0\\
0&0&0&0&0&-1&0&0&0\\
1&0&0&0&0&0&0&0&0\\
0&-1&0&0&0&0&0&0&0\\
0&0&-1&0&0&0&0&0&0\\
0&0&0&0&0&0&1&0&0\\
0&0&0&0&0&0&0&-1&0\\
0&0&0&0&0&0&0&0&-1
\end{pmatrix},
\label{eq:C2}
\end{eqnarray}
\begin{eqnarray}
{\Gamma}(C_{3}^{1})=
\begin{pmatrix}
0&0&0&-\frac{1}{2}&-\frac{\sqrt{3}}{2}&0&0&0&0\\
0&0&0&\frac{\sqrt{3}}{2}&-\frac{1}{2}&0&0&0&0\\
0&0&0&0&0&1&0&0&0\\
0&0&0&0&0&0&-\frac{1}{2}&-\frac{\sqrt{3}}{2}&0\\
0&0&0&0&0&0&\frac{\sqrt{3}}{2}&-\frac{1}{2}&0\\
0&0&0&0&0&0&0&0&1\\
-\frac{1}{2}&-\frac{\sqrt{3}}{2}&0&0&0&0&0&0&0\\
\frac{\sqrt{3}}{2}&-\frac{1}{2}&0&0&0&0&0&0&0\\
0&0&1&0&0&0&0&0&0
\end{pmatrix}
\label{eq:C3}
\end{eqnarray}
Invariance under action of the $C_{3v}$ symmetry group imposes 
\begin{eqnarray}
\left\{\begin{array}{ll}
{\Gamma}(\sigma_{v}^{2})\,\mathcal{\hat{J}}\,{\Gamma}^{t}(\sigma_{v}^{2})=\mathcal{\hat{J}}\\
{\Gamma}(C_{3}^{1})\,\mathcal{\hat{J}}\,{\Gamma}^{t}(C_{3}^{1})=\mathcal{\hat{J}}
\end{array} \right. .
\label{eq:Jinvariant}
\end{eqnarray}
Out of the initial 27 coupling parameters, only six remain independent after imposing Eq.~(\ref{eq:Jinvariant}). The remaining coupling parameters are most elegantly presented in the $\hat{J}_{01}$ coupling matrix
\begin{align}
\label{eq:J01gen}
\hat{J}_{01} = \begin{pmatrix}
J_x & D_z & D_y \\
-D_z & J_y & K\\
-D_y & K & J_z
\end{pmatrix},
\end{align}
the other two coupling matrices taking the form
\begin{eqnarray}
\label{eq:J12gen}
&&\hat{J}_{12} = \nonumber \\
&&\begin{pmatrix}
\frac{1}{4} \left( J_x + 3 J_y \right) & \frac{\sqrt{3}}{4} \left( J_x - J_y \right) + D_z 
& \frac{1}{2} \left( -D_y + \sqrt{3} K \right) \\
\frac{\sqrt{3}}{4} \left( J_x - J_y \right) - D_z & \frac{1}{4} \left( 3 J_x + J_y \right) 
& \frac{1}{2} \left( -\sqrt{3}D_y -  K \right)  \\
\frac{1}{2} \left(    D_y +\sqrt{3}K \right)  & \frac{1}{2} \left( \sqrt{3}D_y -  K \right)  & J_z
\end{pmatrix} \nonumber \\
\end{eqnarray}
\begin{eqnarray}
\label{eq:J20gen}
&&\hat{J}_{20} = \nonumber \\
&&\begin{pmatrix}
\frac{1}{4} \left( J_x + 3 J_y \right) & \frac{\sqrt{3}}{4} \left( J_y- J_x \right) + D_z 
& \frac{1}{2} \left( -D_y - \sqrt{3} K \right)  \\
\frac{\sqrt{3}}{4} \left( J_y - J_x \right) - D_z & \frac{1}{4} \left( 3 J_x + J_y \right) & 
\frac{1}{2} \left( \sqrt{3} D_y -  K \right) \\
\frac{1}{2} \left( D_y -\sqrt{3} K \right) & 
\frac{1}{2} \left(-\sqrt{3}  D_y - K \right) & J_z
\end{pmatrix}. \nonumber \\
\end{eqnarray}

Eqs. (\ref{eq:J01gen})-(\ref{eq:J20gen}) define the 
symmetry allowed nearest neighbour
couplings respecting the $C_{3v}$ point group symmetries of the kagome lattice.\\

For the remainder of this manuscript we will consider the high-symmetry model, with the additional symmetry that the kagome plane itself is a mirror plane of the system. This symmetry is represented by the action of the matrix
\begin{eqnarray}
\Gamma(\sigma_h)=
\begin{pmatrix}
-1&0&0&0&0&0&0&0&0\\
0&-1&0&0&0&0&0&0&0\\
0&0&1&0&0&0&0&0&0\\
0&0&0&-1&0&0&0&0&0\\
0&0&0&0&-1&0&0&0&0\\
0&0&0&0&0&1&0&0&0\\
0&0&0&0&0&0&-1&0&0\\
0&0&0&0&0&0&0&-1&0\\
0&0&0&0&0&0&0&0&1
\end{pmatrix},
\label{eqsigmah}
\end{eqnarray}
on $\hat{\mathcal{J}}$. Constraining the exchange matrices in Eq. (\ref{eq:J01gen})-(\ref{eq:J20gen}) to be invariant
under this additional symmetry, 
\begin{eqnarray}
{\Gamma}(\sigma_{h})\,\mathcal{\hat{J}}\,{\Gamma}^{t}(\sigma_{h})=\mathcal{\hat{J}}
\label{eq:Jsigmah}
\end{eqnarray}
we obtain
\begin{eqnarray}
\left\{\begin{array}{ll}
D_y=0\\
K=0
\end{array}\right.
\label{eq:DK}
\end{eqnarray}
The coupling matrices then become
\begin{align}
\label{eq:J01}
\hat{J}_{01} = 
\begin{pmatrix}
J_x & D_z & 0 \\
-D_z & J_y & 0\\
0 & 0 & J_z
\end{pmatrix},
\end{align}
\begin{align}
\label{eq:J12}
\hat{J}_{12} = \begin{pmatrix}
\frac{1}{4} \left( J_x + 3 J_y \right) & \frac{\sqrt{3}}{4} \left( J_x - J_y \right) + D_z 
& 0 \\
\frac{\sqrt{3}}{4} \left( J_x - J_y \right) - D_z & \frac{1}{4} \left( 3 J_x + J_y \right) 
&0 \\
0 & 0  & J_z
\end{pmatrix}
\end{align}
\begin{align}
\label{eq:J20}
\hat{J}_{20} = \begin{pmatrix}
\frac{1}{4} \left( J_x + 3 J_y \right) & \frac{\sqrt{3}}{4} \left( J_y- J_x \right) + D_z 
& 0\\
\frac{\sqrt{3}}{4} \left( J_y - J_x \right) - D_z & \frac{1}{4} \left( 3 J_x + J_y \right) & 
0\\
0 & 
0 & J_z
\end{pmatrix}.
\end{align}

%======================================================
%======================================================
\subsection{Local XYZ model with\\out-of-plane Dzyaloshinskii-Moriya}
\label{sec:XYZDMham}

The generic nearest-neighbour Hamiltonian on kagome can be written as follows
\begin{align}
\mathcal{H}=\sum_{\Delta} \sum_{\langle ij\rangle}^{} \sum_{\alpha, \beta}^{} S_i^\alpha \hat{J}_{ij}^{\alpha \beta} S_j^\beta,
\label{eq:Hamiltonian}
\end{align}
where the sums are made on all triangles $\Delta$, between nearest neighbours $\langle ij\rangle$, and over all spin components $\alpha,\beta\in\{x,y,z\}$. For the most general case, respecting the $C_{3v}$ point group symmetry, the coupling matrices $\hat{J}_{ij}$ are given in Eqs.~(\ref{eq:J01gen}-\ref{eq:J20gen}). In presence of an additional mirror symmetry in the kagome plane, $\hat{J}_{ij}$ takes the form of Eqs.~(\ref{eq:J01}-\ref{eq:J20}).

%======================
\begin{figure}[h]
\includegraphics[scale=1]{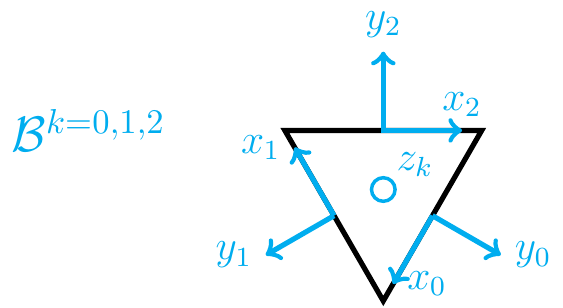}
\caption{
Local bases $\mathcal{B}^{k}$ used in Eq.~(\ref{eq:JbasisB}) where the generic nearest-neighbour Hamiltonian on kagome takes the form of an XYZ model with Dzyaloshinskii-Moriya (DM) interactions. All $z_{k}$ axes are pointing out of plane. All $x_{k}$ ($y_{k}$) axes are pointing along (orthogonal to) their local bond.}
\label{fig:basisB}
\end{figure}
%======================

A consequence of the mirror symmetry of the kagome plane is the decoupling between in-plane $\mathbf{S}_{i}^{\perp}$ and out-of-plane $S_{i}^{z}$ spin components [Eq.~(\ref{eq:DK})]. In materials where the kagome layer is embedded in a three-dimensional structure, this mirror symmetry can be broken by surrounding ions. This is the case for example in Jarosites~\cite{Elhajal2002,Chernyshev15b,Chernyshev15a} and tripod kagome materials \cite{Dun16a,Scheie16a,Paddison16b,Dun17a}. Here, we focus on models respecting the full kagome symmetry. This means among other things that in-plane Dzyaloshinskii-Moriya couplings are forbidden~\cite{Elhajal2002}. But out-of-plane ones are not. In the coupling matrices $\hat{J}_{ij}$, out-of-plane DM interactions are parametrised by the antisymmetric term $D_{z}$, whose traditional form in a Hamiltonian is
\begin{align}
\mathcal{H}_{\rm DM} = {\bf D} \cdot ({\bf S}_i \times {\bf S}_j)
\textrm{ with  }{\bf D} = (0,0,D_{z}).
\label{eq:DM}
\end{align}
From now on, we shall simply write $D=D_{z}$.

The expression of the coupling matrices in Eqs.~(\ref{eq:J12}-\ref{eq:J20}) is not necessarily very insightful. In the appropriate, bond-dependent, local bases $\mathcal{B}^{k}$ given in Fig.~\ref{fig:basisB}, one can take advantage of the kagome symmetry by $\pm 2\pi/3$ rotations to rewrite the three $\hat{J}_{ij}$ matrices in the same, more convenient, form
\begin{eqnarray}
\hat{J}_{01}^{B^{2}} = 
\hat{J}_{12}^{B^{0}} = 
\hat{J}_{20}^{B^{1}} = 
\begin{pmatrix}
J_x & D & 0 \\
-D & J_y & 0 \\
0 & 0 & J_z
\end{pmatrix}.
\label{eq:JbasisB}
\end{eqnarray}

As a summary, the generic nearest-neighbour Hamiltonian respecting the full symmetry of the kagome lattice is a local XYZ model with out-of-plane Dzyaloshinskii-Moriya (DM) interactions [Eq.~(\ref{eq:JbasisB})]. We label this model XYZDM. One should emphasise that it is not a traditional XYZ model, as would be the case if it was expressed in the same global frame for all bonds. Such a global XYZ Hamiltonian is not allowed by the symmetry of the kagome lattice, assuming that the spins transform as axial vectors under the point group operations.

%======================================================
%======================================================
\subsection{Related generic models}
\label{sec:related}

Among the related generic systems that have been studied in the literature, one should mention the generic quantum spin Hamiltonian on the triangular lattice\cite{Li15a,Li16a,Luo17a,Li16c}, quantum kagome ice \cite{Carrasquilla15a}, the spin-orbital liquids of non-Kramers magnets \cite{Schaffer13a} and the classical regular-magnetic-order classification of Messio \textit{et al.} \cite{Messio2011}.

The former is the triangular version of the present kagome Hamiltonian, which has been particularly successful in describing the spin liquid candidate YbMgGaO$_{4}$ \cite{Li15a,Paddison17a}. Besides the obvious fact that triangular and kagome lattices are different, one of the main distinctions between the two microscopic Hamiltonians is the absence of Dzyaloshinskii-Moriya interactions in triangular systems because of inversion symmetry. The propensity of the DM coupling to induce an intrinsic magnetic chirality will be a recurrent feature of our work.

Quantum kagome ice has been studied in the context of a pyrochlore lattice in a strong [111] magnetic field \cite{Carrasquilla15a}, where the ``spin'' degrees of freedom correspond to the states of a ``dipole-octupole'' crystal-field doublet \cite{Huang14a}. It is thus inherently different from the generic model studied here, but remains a motivation for future applications of our work, in particular the inclusion of quantum fluctuations. Quantum kagome ice is indeed a promising candidate for a gapped $\mathbb{Z}_{2}$ spin liquid ground-state \cite{Carrasquilla15a,Owerre16b,Owerre16ab,Huang17a}, where disclination defects have been proposed to host symmetry-protected vison zero modes \cite{Huang17a}.

Furthermore, concerning the inclusion of quantum fluctuations, a projective symmetry group analysis has investigated possible spin-orbital liquids with fermionic spinons for non-Kramers pseudospin$-1/2$ models \cite{Schaffer13a}. The unusual time-reversal symmetry of non-Kramers ions steps away from our present study, but is an interesting aspect of generic models \cite{Onoda11a,Schaffer13a}, that has been shown to support magnetic phases forbidden for Kramers pseudospin$-1/2$ kagome models.

As for the classification of Ref.~[\onlinecite{Messio2011}], it is a group theoretical approach, applied to a variety of lattices including kagome, 
able to list all regular magnetic orders which respect the lattice symmetries modulo global O(3) spin transformations. It has been used \textit{e.g.} in studying the candidate quantum spin liquid material, Kapellasite \cite{Fak2012}. Even if the lattice symmetry plays a key role in both our approaches, the constraint of global O(3) symmetry prevents the 
consideration of most models with anisotropic interactions, which represents the ``bulk'' of the XYZDM model. 
Nevertheless, the regular magnetic orders will reappear in our work for Hamiltonians tuned 
precisely on high-symmetry points, where a global O(3) invariance reappears.

%======================================================
%======================================================
%======================================================
\section{Hamiltonian diagonalization}
\label{sec:irrep}

Now that our model has been determined, let us explore the phases it begets. In this section \ref{sec:irrep}, we will see how the irrep decomposition provides the eigenbasis of order parameters necessary to diagonalize the coupling matrix $\mathcal{\hat J}$. The general method to determine the ground-states is exposed in section \ref{sec:howGS}. We refer the reader to Ref.~[\onlinecite{Yan17a}] to see this method applied to pyrochlores.

%======================================================
%======================================================
\subsection{Irreducible Representations}
\label{sec:def}

Any spin configuration on a triangle can be described by the $9-$dimensional vector of Eq.~(\ref{eq:repres}). Let $\Gamma(g)$ be the $9\times 9$ matrix representing the element $g\in C_{3v}= \left\{e, C_3^{i=1,2}, \sigma_v^{i=0,1,2}\right\}$ in the global Cartesian basis, as exemplified in Eqs.~(\ref{eq:C2}) and (\ref{eq:C3}). By definition, the $\Gamma$ matrices provide a $9-$dimensional representation of the $C_{3v}$ symmetry group.

The $\Gamma$ representation is said to be reducible if there is a unitary transformation $\hat U$ such that $\hat U \,\Gamma(g)\, \hat U^{-1}$ is block-diagonal, with the same block structure, for all $g\in C_{3v}$. If the blocks cannot be further reduced, \textit{i.e.} if they are ``as small as possible'', then each block is an irreducible representation (irrep) of $C_{3v}$ in its own subspace. The interest of such an irreducible decomposition is that it is valid for any matrix invariant under action of the $C_{3v}$ symmetry group. Once rewritten in the basis provided by  $\hat U$, $\mathcal{\hat J}$ is greatly simplified as it can only couple basis vectors transforming according to the same irrep.
This method brings us a stone's throw from the full diagonalisation of the Hamiltonian.

The decomposition of the $\Gamma$ representation can be formally written as a direct sum of the irreps $\Gamma_{I}$
\begin{align}
\Gamma &= \bigoplus_{I} \gamma_I \Gamma_{I},
\label{eq:Reducible-Representation}
\end{align}
where each irrep $\Gamma_{I}$ appears $\gamma_I$ times in the decomposition. For any symmetry operations $g\in C_{3v}$, the trace of $\Gamma_{I}(g)$ is called its character, $\chi_{I}(g)$. The character of $\Gamma(g)$ is $\chi(g)$. In terms of these characters, Eq.~(\ref{eq:Reducible-Representation}) translates to
\begin{align}
\chi(g) = \sum_{I} \gamma_I \chi_{I}(g) {\,} , {\qquad} \forall g \in C_{3v}.
\end{align}
The coefficients $\gamma_I$ can be found using the formula \cite{Tinkham64a}
\begin{align}
\gamma_I = \frac{1}{n} \sum_{g \in C_{3v}} \chi_{I}(g) \chi(g),
\label{eq:gamma}
\end{align}
where $n$ is the order of the group ($n=6$). The irreps and character of the $C_{3v}$ symmetry group are tabulated, and can be read in Appendix B of Ref.~[\onlinecite{Tinkham64a}] for example. As for $\chi(g)$, it is directly obtained from Eqs.~(\ref{eq:C2}) and (\ref{eq:C3}), the trace of the neutral element $e$ being trivial. All characters are summarized in table \ref{tab:Character-Table}
\begin{table}[h!]
\large
\begin{tabular}{c|l*{3}{c}}
$C_{3v}$ & e & $C_3$ & $\sigma_{v}$ \\
\hline
$\Gamma_{1} = A_1$ & 1 & 1 & 1 \\
$\Gamma_{2} = A_2$ & 1 & 1 & -1 \\
$\Gamma_{3} = E$ & 2 & -1 & 0	\\
\hline
$\Gamma$ & 9 & 0 & -1	
\end{tabular}
\caption{Character table of the point group $C_{3v}$. $e$, $C_3$ and $\sigma_v$ correspond to the three conjugacy classes of $C_{3v}$. We have used the Mulliken symbols for the notation of the irreducible representations $\Gamma_{I}$. The last line corresponds to the characters of the reducible representation $\Gamma$.}
\label{tab:Character-Table}
\end{table}\\
Using Eq.~(\ref{eq:gamma}) and table \ref{tab:Character-Table}, we find
\begin{align}
\label{eq:Representation-Decomposition}
\gamma_1 = 1, \gamma_2 = 2 \text{ and } \gamma_3 = 3.
\end{align}
In practice, it means that the coupling matrix $\mathcal{\hat J}$ of Eq.~(\ref{eq:9x9}) can be block-diagonalized into 6 blocks: $\gamma_{1}+\gamma_{2}=3$ scalar blocks (corresponding to $A_{1}$ and $A_{2}$) and $\gamma_{3}=3$ blocks of size $2\times 2$  (corresponding to $E$).

%======================================================
%======================================================
\subsection{Basis Vectors}
\label{sec:bv}

The basis vectors of this block-diagonalization must obey the same symmetry properties as the irreps they correspond to (for more details, see Ref. [\onlinecite{Dresselhaus2008}]). But how to calculate these basis vectors? From the coupling matrices in Eqs.(\ref{eq:J01}) - (\ref{eq:J20}), the $xy$-components are decoupled from the $z$-component. Hence, an appropriate choice of basis should not mix the in-plane and out-of-plane components of the spins. This translates to 6 (resp. 3) basis vectors with only in-plane (resp. out-of-plane) spin components.

\begin{figure}[b]
\centering
\hspace{0.4cm}\includegraphics[scale=0.75]{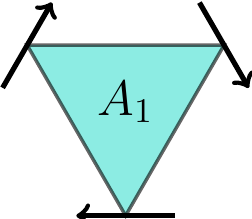}\\
\hspace{0.4cm}\includegraphics[scale=.75]{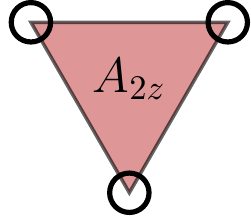}
\hspace{2.4cm}\includegraphics[scale=.75]{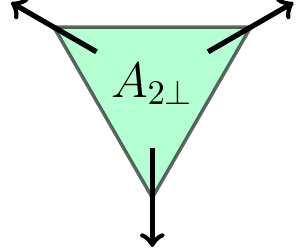}\\
\includegraphics[scale=.75]{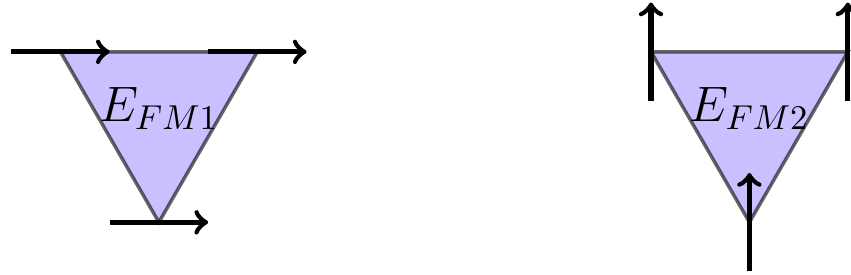}\\
\hspace{0.5cm}\includegraphics[scale=.75]{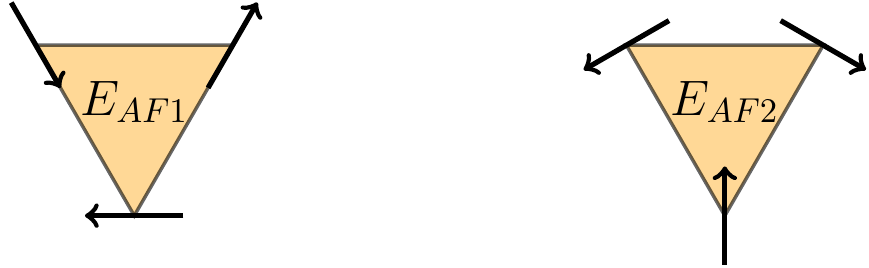}
\caption{Spin configurations corresponding to different irreducible representations, as expressed in Eqs.~(\ref{eq:A1}-\ref{eq:Eb}). $E_{FM1}, E_{FM2}$ and $A_{2z}$ have saturated magnetization respectively along the $x, y$ and $z$ axes. $A_{1}$ and $A_{2\perp}$ have maximum negative vector chirality, while $E_{AF}$ has maximum positive vector chirality [Eq.~(\ref{eq:chivec})]. When considered together, the $A_{1}$ and $A_{2\perp}$ irreps are labeled $A$ for convenience.
}
\label{fig:irrep}
\end{figure}

For one-dimensional representations, the group elements are scalars and they reduce to the character itself. In the trivial one-dimensional irreducible representation $A_1$, all the elements are equal to one. Hence, we are looking for a state invariant under all $C_{3v}$ symmetries. The only possibility is:
\begin{align}
\tilde{S}(A_1) = 
 \left(\frac{1}{2}, \frac{\sqrt{3}}{2}, 0, \frac{1}{2}, -\frac{\sqrt{3}}{2}, 0, -1, 0, 0 \right),
\label{eq:A1}
\end{align}
illustrated in Fig.~\ref{fig:irrep}. Please keep in mind that the spins transform like axial vectors, \textit{i.e.} they are invariant under lattice inversion [Eq.~(\ref{eq:SinvS})].

%======================
%======================

The irrep $A_2$ appears twice in $\Gamma$ [Eq.(\ref{eq:Representation-Decomposition})]. Thus we need two basis vectors for this representation. Since a $A_2$ irrep corresponds again to a one-dimensional representation, it is easy to see how the elements act on the vectors. According to the character table ({\it cf.} table \ref{tab:Character-Table}), we are looking for two vectors $\tilde{V}_{i=1,2}$ that are invariant under $C_3$ and change from $\tilde{V}_{i}$ to $-\tilde{V}_{i}$ under $\sigma_v$. The only solutions are:
\begin{align}
\tilde{S}_{z}(A_2) = (0, 0, 1, 0, 0, 1, 0, 0, 1)
\label{eq:A2a}
\end{align}
and
\begin{align}
\tilde{S}_{\perp}(A_2) =
 \left(-\frac{\sqrt{3}}{2}, \frac{1}{2}, 0, \frac{\sqrt{3}}{2}, \frac{1}{2}, 0, 0, -1, 0 \right)
\end{align}

The last irrep $E$ is of dimension $2$ and appears $3$ times in $\Gamma$. Therefore one needs to find $3$ different pairs of basis vectors for this representation. Since the group elements do not reduce to their character anymore, this is less straightforward than for the $A$ irreps. Also, the choice of basis vectors is not unique, but can be made physically intuitive. By definition of the irrep decomposition, each pair of basis vectors shall generate an invariant subspace under action of the $C_{3v}$ symmetry group. All elements of the $C_{3v}$ symmetry group can be described as a successive permutation of the spin positions and global rotation of the spin orientations. Such transformations trivially conserves the norm of the total magnetic moment. It means that the subspace of saturated configurations, \textit{i.e.} with collinear spins, is invariant under action of the $C_{3v}$ symmetry group. Magnetisation along the $z$-axis has already been accounted for by the $A_{2z}$ basis vector [Eq.~(\ref{eq:A2a})]. We are thus left with the subspace of configurations with saturated in-plane magnetization which, by decoupling of the $xy$ and $z$ spin components previously mentioned, is also invariant under action of all $C_{3v}$ elements. This subspace thus corresponds to an $E$ irrep, labeled $E_{FM}$. A natural choice of basis for $E_{FM}$ is
\begin{align}
\left\{\begin{array}{ll}
\tilde{S}_{1, FM} (E) &= (1,0,0,1,0,0,1,0,0)  \\
\tilde{S}_{2, FM} (E) &= (0,1,0,0,1,0,0,1,0)
\end{array} \right. ,
\end{align}
as depicted in Fig.~\ref{fig:irrep}.

Hence, out of the six expected basis vectors with in-plane spin components for the representation $\Gamma$ of Eq.~(\ref{eq:repres}), four of them have now been determined, namely $A_{1}$, $A_{2\perp}$ and $E_{FM}$. By imposing the orthogonality of the basis, the remaining two in-plane basis vectors are given in Eq.~(\ref{eq:Eb}). The subspace they generate is invariant under action of the $C_{3v}$ symmetry group; the corresponding irrep is labeled $E_{AF}$ and represented in Fig.~\ref{fig:irrep}.
\begin{align}
\left\{\begin{array}{ll}
\tilde{S}_{1, AF} (E) &= \left(\frac{1}{2},-\frac{\sqrt{3}}{2},0,\frac{1}{2},\frac{\sqrt{3}}{2},0,-1,0,0 \right)  \\
\tilde{S}_{2, AF} (E) &= \left(-\frac{\sqrt{3}}{2},-\frac{1}{2},0,\frac{\sqrt{3}}{2},-\frac{1}{2},0,0,1,0 \right)
\end{array} \right.
\label{eq:Eb}
\end{align}
Within the $E_{FM}$ and $E_{AF}$ subspaces, the basis vectors are orthogonal for each sublattice $i\in\{0,1,2\}$:
\begin{align}
\left\{\begin{array}{ll}
	\sum\limits_{\alpha=1}^{3} \left(\tilde{S}_{1,FM}(E)\right)_{3i+\alpha} \cdot \left(\tilde{S}_{2,FM}(E)\right)_{3i+\alpha} &= 0 \\
	\\
	\sum\limits_{\alpha=1}^{3} \left(\tilde{S}_{1,AF}(E)\right)_{3i+\alpha} \cdot \left(\tilde{S}_{2,AF}(E)\right)_{3i+\alpha} &= 0 
\end{array} \right.
\end{align}

The last two basis vectors correspond to antiferromagnetic states with out-of-plane spin components. The corresponding subspace is invariant under action of the $C_{3v}$ symmetry group and is labeled $E_{z}$. A possible choice of basis for this subspace is
\begin{align}
\left\{\begin{array}{ll}
\tilde{S}_{1z} (E) &= \sqrt{\frac{3}{2}}\left(0,0,1,0,0,-1,0,0,0 \right)  \\
\tilde{S}_{2z} (E) &= \frac{1}{\sqrt{2}}\left(0,0,1,0,0,1,0,0,-2 \right)
\end{array} \right. ,
\label{eq:EcBasis}
\end{align}
whose spins are not normalized
\begin{align}
\left\{\begin{array}{ll}
\sum\limits_{\alpha=1}^{3} \left(\tilde{S}_{1z}(E)\right)_{3i+\alpha}^2 &\neq 1 \\
\\
\sum\limits_{\alpha=1}^{3} \left(\tilde{S}_{2z}(E)\right)_{3i+\alpha}^2 &\neq 1 
\end{array} \right. .
\label{eq:E-non-normal}
\end{align}
The inequalities (\ref{eq:E-non-normal}) are not a consequence of the particular choice of basis in Eq. (\ref{eq:EcBasis}). Within the subspace generated by $\tilde{S}_{1z}(E)$ and $\tilde{S}_{2z}(E)$, it is impossible to find a configuration where the three spins are all normalized. The reason is trivially because it is impossible for three normalized collinear spins to bear zero magnetization. This is an important property of the kagome lattice, which will be discussed in detail throughout the paper.
%======================================================
%======================================================
\subsection{Order parameters}
\label{ssec:op}

The vector $\tilde{S}$ can be expressed in terms of the irreps basis:
\begin{align}
\begin{split}
\tilde{S} &= {\,} m_{A_1} \tilde{S}(A_1) + m_{A_2,z} \tilde{S}_{z}(A_2) + m_{A_2,\perp} \tilde{S}_{\perp}(A_2) \\
&+ m_{E,FM}^x \tilde{S}_{1,FM}(E) + m_{E,FM}^y \tilde{S}_{2,FM}(E) + m_{E,AF}^x \tilde{S}_{1,AF}(E)\\
& + m_{E,AF}^y \tilde{S}_{2,AF}(E) + m_{E,z}^x \tilde{S}_{1z}(E) + m_{E,z}^y \tilde{S}_{2z}(E)
\end{split}
\label{eq:Sop}
\end{align}
with
\begin{align}
m_{\alpha,i} = \frac{1}{3} \tilde{S} \cdot \tilde{S}_i(\alpha)
\end{align}
being the order parameters associated with the irreducible representations:
\begin{align}
\label{eq:Order-Parameter-A1}
m_{A_1} &= \frac{1}{3} \left( \frac{1}{2} S_0^x + \frac{\sqrt{3}}{2} S_0^y + \frac{1}{2} S_1^x - \frac{\sqrt{3}}{2} S_1^y - S_2^x \right)\\
\label{eq:Order-Parameter-A2a}
m_{A_2,z} &= \frac{1}{3} \left( S_0^z + S_1^z + S_2^z \right)\\
\label{eq:Order-Parameter-A2b}
m_{A_2,\perp} &= 
\frac{1}{3} \left( -\frac{\sqrt{3}}{2} S_0^x + \frac{1}{2} S_0^y + \frac{\sqrt{3}}{2} S_1^x + \frac{1}{2} S_1^y - S_2^y \right)\\
\label{eq:Order-Parameter-Ea}
{\bf m}_{E, FM} &= \frac{1}{3} \begin{pmatrix}
S_0^x + S_1^x + S_2^x \\
S_0^y + S_1^y + S_2^y
\end{pmatrix}\\
\label{eq:Order-Parameter-Eb}
{\bf m}_{E, AF} &= \frac{1}{3} \begin{pmatrix}
\frac{1}{2} S_0^x - \frac{\sqrt{3}}{2} S_0^y + \frac{1}{2} S_1^x + \frac{\sqrt{3}}{2} S_1^y - S_2^x \\
-\frac{\sqrt{3}}{2} S_0^x - \frac{1}{2} S_0^y + \frac{\sqrt{3}}{2} S_1^x - \frac{1}{2} S_1^y + S_2^y
\end{pmatrix}\\
\label{eq:Order-Parameter-Ec}
{\bf m}_{E, z} &= \frac{1}{3} \begin{pmatrix}
\sqrt{\frac{3}{2}}(S_0^z - S_1^z) \\
\frac{1}{\sqrt{2}} (S_0^z + S_1^z - 2 S_2^z)
\end{pmatrix}
\end{align}
By decomposition of Eq.~(\ref{eq:Sop}), the order parameters obey the relation
\begin{align}
\label{eq:m1}
\begin{split}
& m_{A_1}^2 + m_{A_2,z}^2 + m_{A_2,\perp}^2 + {\bf m}_{E, FM}^2 + {\bf m}_{E, AF}^2 + {\bf m}_{E, z}^2 \\
&= \frac{1}{3} \left({\bf S}_0^{\,2} + {\bf S}_1^{\,2} + {\bf S}_2^{\,2} \right) = 1
\end{split}
\end{align}
Please note that $\max(\left|{\bf m}_{E,z}\right|^2) < 1$, which means that order into the $E_{z}$ phase necessarily co-exists with other phases.\\

Alternatively, one can write the spin configurations as a function of the order parameters:
\begin{align}
\label{eq:S0}
&\left\{\begin{array}{ll}
S_0^x &= \frac{1}{2} m_{A_1} - \frac{\sqrt{3}}{2} m_{A_2,\perp} + m^{x}_{E, FM} + \frac{1}{2} m^{x}_{E, AF} \\
&- \frac{\sqrt{3}}{2} m^{y}_{E, AF} \\
S_0^y &= \frac{\sqrt{3}}{2} m_{A_1} + \frac{1}{2} m_{A_2,\perp} + m^{y}_{E, FM} - \frac{\sqrt{3}}{2} m^{x}_{E, AF} \\
&- \frac{1}{2} m^{y}_{E, AF} \\
S_0^z &= m_{A_2,z} + \sqrt{\frac{3}{2}} m^{x}_{E, z} + \frac{1}{\sqrt{2}} m^{y}_{E, z}
\end{array}\right.\\
\label{eq:S1}
&\left\{\begin{array}{ll}
S_1^x &= \frac{1}{2} m_{A_1} + \frac{\sqrt{3}}{2} m_{A_2,\perp} + m^{x}_{E, FM} + \frac{1}{2} m^{x}_{E, AF} \\
&+ \frac{\sqrt{3}}{2} m^{y}_{E, AF} \\
S_1^y &= -\frac{\sqrt{3}}{2} m_{A_1} + \frac{1}{2} m_{A_2,\perp} + m^{y}_{E, FM} + \frac{\sqrt{3}}{2} m^{x}_{E, AF} \\
&- \frac{1}{2} m^{y}_{E, AF} \\
S_1^z &= m_{A_2,z} - \sqrt{\frac{3}{2}} m^{x}_{E, z} + \frac{1}{\sqrt{2}} m^{y}_{E, z}
\end{array}\right.\\
\label{eq:S2}
&\left\{\begin{array}{ll}
S_2^x &= - m_{A_1} + m^{x}_{E, FM} - m^{x}_{E, AF} \\
S_2^y &= - m_{A_2,\perp} + m^{y}_{E, FM} + m^{y}_{E, AF} \\
S_2^z &= m_{A_2,z} - \sqrt{2} m^{y}_{E, z}
\end{array}\right.
\end{align}

%======================================================
%======================================================
\subsection{Hamiltonian in quadratic form}
\label{sec:diagham}

In terms of the order parameters, the Hamiltonian of Eq.~(\ref{eq:Hamiltonian}) can be rewritten as
\begin{align}
\label{eq:ham}
\mathcal{H} = \sum_{\Delta} \frac{3}{2}
(&\lambda_{A_1} m_{A_1}^2 + \lambda_{A_2, z} m_{A_2,z}^2 + \lambda_{A_2, \perp} m_{A_2,\perp}^2 \nonumber\\
+&\lambda_{E, FM}\,{\bf m}_{E, FM}^2 + \lambda_{E, AF}\,{\bf m}_{E, AF}^2 + \lambda_{E, z}\,{\bf m}_{E, z}^2 \nonumber\\
+&\left.\lambda_{E, mix} {\,} {\bf m}_{E, FM} {\,} \cdot {\bf m}_{E, AF} \right)
\end{align}
where the sum is over all triangles $\Delta$ in the kagome lattice, and the coefficients $\lambda_i$ are
\begin{align}
\label{eq:lambda-A1}
&\lambda_{A_1} = \frac{1}{2} \left(J_x - 3 J_y - 2 \sqrt{3} D \right) \\
\label{eq:lambda-A2a}
&\lambda_{A_2, z} = 2 J_{z} \\
\label{eq:lambda-A2b}
&\lambda_{A_2, \perp} = \frac{1}{2} \left( -3 J_x + J_y  - 2 \sqrt{3} D \right) \\
\label{eq:lambda-Ea}
&\lambda_{E, FM} = J_x + J_y \\
\label{eq:lambda-Eb}
&\lambda_{E, AF} = \frac{1}{2} \left( - J_x - J_y + 2 \sqrt{3} D \right) \\
\label{eq:lambda-Eab}
&\lambda_{E, mix} = J_x - J_y \\
\label{eq:lambda-Ec}
&\lambda_{E, z} = -J_z
\end{align}
To avoid any confusion, one should probably insist that the Hamiltonian of Eq.~(\ref{eq:ham}) is \textit{not} a Landau mean-field expansion, but an exact rewriting of the original Hamiltonian of Eq.~(\ref{eq:Hamiltonian}). The present decomposition is the final outcome of the block diagonalization of sections \ref{sec:def} and \ref{sec:bv}. Hence, it prevents the mixing between inequivalent irreps. In the absence of the reflection symmetry in the kagome plane [Eq. (\ref{eq:Jsigmah})] there would be allowed couplings between $A_{2z}$ and $A_{2\perp}$ on one hand, and $E_{FM}, E_{AF}$ and $E_{z}$ on the other hand. Once this symmetry is imposed, however, there is no coupling between the $xy$-plane and the $z$-axis, the only possible coupling term is between $E_{FM}$ and $E_{AF}$. This coupling term is coming from our physically intuitive, but mathematically arbitrary choice of $E_{FM}$ and $E_{AF}$. It can be eliminated with a different choice of basis vectors, whose corresponding order parameters are
\begin{align}
\label{eq:malpha-mbeta}
\begin{split}
{\bf m}_{E,\alpha} &= \cos \phi {\,} {\bf m}_{E, FM} - \sin \phi {\,} {\bf m}_{E, AF} \\
{\bf m}_{E,\beta} &= \sin \phi {\,} {\bf m}_{E, FM} + \cos \phi {\,} {\bf m}_{E, AF}
\end{split}
\end{align}
where $\phi$ is given by
\begin{align}
\phi = \frac{1}{2} \arctan \left(\frac{J_y -  J_x}{\frac{3}{2} (J_x + J_y) - \sqrt{3} D} \right).
\label{eq:phi}
\end{align}
In this basis, the Hamiltonian is now fully quadratic for each triangle
\begin{align}
	\label{eq:hamQ}
	\begin{split}
	& \mathcal{H} = \sum_{\Delta}^{} \frac{3}{2} \left( \lambda_{A_1} m_{A_1}^2 + \lambda_{A_2, z} m_{A_2,z}^2 + \lambda_{A_2, \perp} m_{A_2,\perp}^2 \right. \\
	&\left. + \lambda_{E,\alpha} {\,} {\bf m}_{E,\alpha}^2 + \lambda_{E,\beta} {\,} {\bf m}_{E,\beta}^2 + \lambda_{E, z} {\,} {\bf m}_{E, z}^2 \right),
	\end{split}
\end{align}
with
\begin{align}
\label{eq:lambda-Ealpha}
	\begin{split}
		\lambda_{E,\alpha} &= (J_x + J_y) \cos^2 \phi \\
		&- \frac{1}{2} (J_x + J_y - 2 \sqrt{3} D) \sin^2 \phi \\
		&- \frac{1}{2} (J_x - J_y) \sin(2 \phi),
	\end{split}\\
\label{eq:lambda-Ebeta}
	\begin{split}
	\lambda_{E,\beta} &= (J_x + J_y) \sin^2 \phi \\
	&- \frac{1}{2} (J_x + J_y - 2 \sqrt{3} D) \cos^2 \phi \\
	&+ \frac{1}{2} (J_x - J_y) \sin(2 \phi).
	\end{split}
\end{align}

However, the two new pairs of basis vectors, $E_{\alpha}$ and $E_{\beta}$, correspond to non-normalized spin configurations, as in the case of $E_{z}$.

%======================================================
%======================================================
\subsection{How to determine the ground-states}
\label{sec:howGS}

The choice of any specific model is defined by its coupling parameters $\{J_{x},J_{y},J_{z},D\}$. In the Hamiltonian of Eq.~(\ref{eq:hamQ}), this choice only appears in the eigenvalues $\lambda_{I}$, while the spin variables are entirely embedded in the quadratic terms -- each term corresponding to a different order parameter $m_{I}$. As was done on the pyrochlore lattice~\cite{Yan17a}, the energy can \textit{a priori} be minimized for each triangle by maximizing the value of the order parameter $m_{I_{0}}$ which has the smallest eigenvalue $\lambda_{I_{0}}=\min\{\lambda_{A_1},\lambda_{A_2, z},\lambda_{A_2, \perp},\lambda_{E,\alpha},\lambda_{E,\beta},\lambda_{E, z}\}$. For a uniform lattice, the eigenvalues $\lambda_{I}$ have the same value for all triangles. It means that the ground-state for a given set of coupling parameters is obtained by paving the entire kagome lattice with configurations which saturate the $I_{0}$ order parameter on every triangle: 
$m_{I_{0}}^{2}=1$ and $m_{I\neq I_{0}}^{2}=0$.

If there is more than one minimum eigenvalue, say $\lambda_{I_{0}}=\lambda_{I'_{0}}=\min\{\lambda_{A_1},\lambda_{A_2, z},\lambda_{A_2, \perp},\lambda_{E,\alpha},\lambda_{E,\beta},\lambda_{E, z}\}$, then both $I_{0}$ and $I'_{0}$ configurations are allowed in the ground-state. Such accidental degeneracy occurs for specific values of parameters at the $T=0$ frontier between $I_{0}$ and $I'_{0}$ phases. These frontiers are sometimes the birth places of spin liquids~\cite{Yan17a}.\\

However, there is an important caveat. The unit-length spin constraint must always be respected, $|\mathbf{S}_{i}|=1$. While this constraint is ensured for any of the $A_{1},A_{2z},A_{2\perp},E_{FM}$ and $E_{AF}$ configurations (see Fig.~\ref{fig:irrep}), it is not the case for the $E_{\alpha}, E_{\beta}$ and $E_{z}$ configurations. 
The corresponding order parameters cannot be saturated for any physical spin configuration
\begin{eqnarray}
&& \max(|{\bf m}_{E,z}|^2) < 1\\
&& \max(|{\bf m}_{I}|^2 \,|\, I\in\{E_\alpha,E_\beta\}) < 1\textrm{ when } J_{x}\neq J_{y}
\end{eqnarray}
If the minimum eigenvalue $\lambda_{I_{0}}$ corresponds to $I_{0}\in\{E_\alpha,E_\beta,E_z\}$, then basis vectors associated with other eigenvalues $\lambda_{I\neq I_{0}} \geq \lambda_{I_{0}}$ need to be included in the ground-state. In terms of the irreps decomposition of Eq.~(\ref{eq:Sop}), it means that the ground-state is most likely described by more than one irrep, and its determination becomes a tedious task.

%======================
\begin{figure*}[t]
\centering
\includegraphics[scale=0.69]{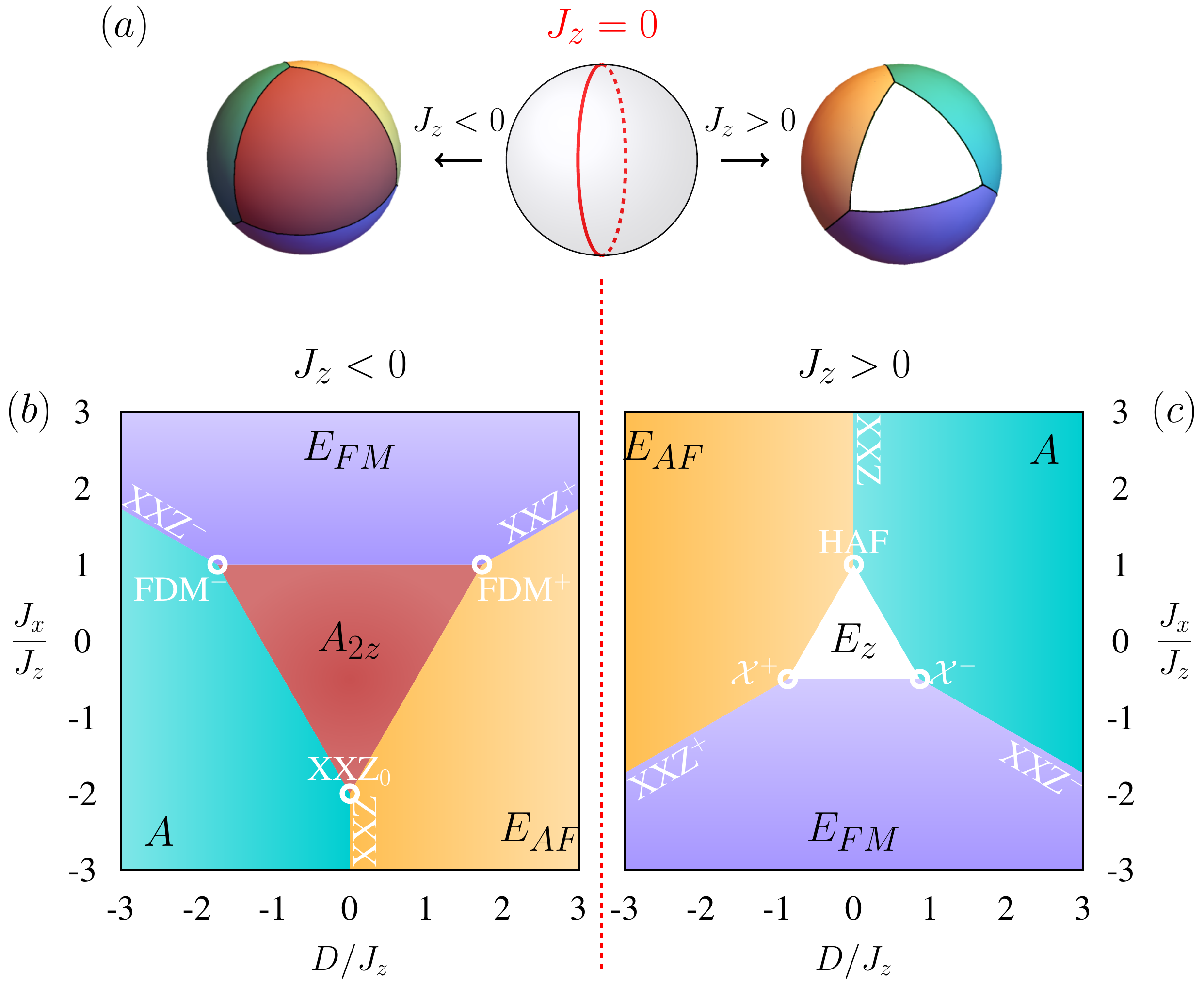}\\\vspace{0.5cm}
\includegraphics[scale=0.8]{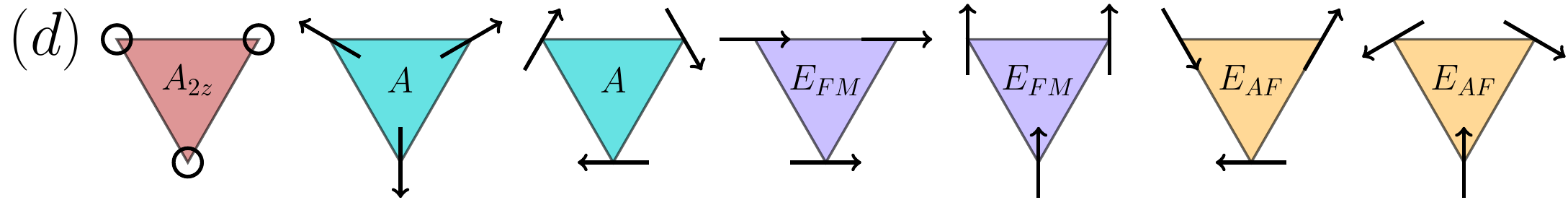}
\caption{
Zero-temperature phase diagram of the XXZ model with Dzyaloshinskii-Moriya interactions (XXZDM). The phase diagram can be divided into five continuous regions when represented on a sphere $(a)$, with spherical coordinates $(\theta,\varphi)$: $(J_{x},D,J_{z})=(\cos\varphi \sin\theta, -\sin\varphi \sin\theta, \cos\theta)$. The three-fold symmetry of the XXZDM kagome model~\cite{Essafi16a} is transparent in this representation. The left/right hemispheres correspond to ferromagnetic/antiferromagnetic coupling $J_{z}$, which can be projected onto two planar phase diagrams for a quantitative comparison, respectively panels $(b)$ and $(c)$. $A_{2z}$ and $E_{z}$ regions take the form of triangles centred at the ``poles'' of each hemisphere. The latter is noticeably smaller because of its comparatively high antiferromagnetic frustration, making it less energetically stable when competing with the surrounding ordered phases. The name of specific models with extensive degeneracy are written in white, as defined in Ref.~[\onlinecite{Essafi16a}] and given in Table \ref{tab:SL}. For convenience the corresponding spin configurations are copied from Fig.~\ref{fig:irrep} in panel $(d)$. The $A$ region corresponds to $A_{1}$ and $A_{2\perp}$, which have the same energy in the XXZDM model.
}
\label{fig:PhD}
\end{figure*}
%======================

%======================================================
%======================================================
%======================================================
\section{XXZ model with Dzyaloshinskii-Moriya}
\label{sec:XXZDM}

The diagonalization of Eq.~(\ref{eq:hamQ}) comes at the cost of introducing two pairs of basis vectors, $E_{\alpha}$ and $E_{\beta}$, with non-normalized spins. To circumvent this problem, we can restrict our analysis at first to the region of the phase diagram where the coupling between $E_{FM}$ and $E_{AF}$ vanishes [Eqs.~(\ref{eq:lambda-Eab})]
\begin{eqnarray}
J_{x}=J_y.
\label{eq:Jxy}
\end{eqnarray}
In this region, the coupling matrices of Eqs.(\ref{eq:J01}-\ref{eq:J20}) become
\begin{align}
\label{eq:couplings}
\hat{J}_{01} = \hat{J}_{12} = \hat{J}_{20} =
\begin{pmatrix}
J_x & D & 0 \\
-D & J_x & 0 \\
0 & 0 & J_z
\end{pmatrix}
\end{align}
which corresponds to the XXZ model with Dzyaloshinskii-Moriya interactions (XXZDM)
\begin{align}
\mathcal{H}
&= \sum_{\langle ij\rangle} J_x (S_i^x S_j^x + S_i^y S_j^y) + J_z S_i^z S_j^z + {\bf D} \cdot ({\bf S}_i \times {\bf S}_j).
\end{align}
In terms of the order parameters, the Hamiltonian of Eq.~(\ref{eq:ham}) becomes
\begin{align}
\label{eq:hamXXZ}
\mathcal{H} = \sum_{\Delta} &\frac{3}{2}
(\lambda_{A_1} m_{A_1}^2 + \lambda_{A_2, z} m_{A_2,z}^2 + \lambda_{A_2, \perp} m_{A_2,\perp}^2 \nonumber\\
+&\left.\lambda_{E, FM}\,{\bf m}_{E, FM}^2 + \lambda_{E, AF}\,{\bf m}_{E, AF}^2 + \lambda_{E, z}\,{\bf m}_{E, z}^2 \right),
\end{align}
with
\begin{align}
\label{eq:lA2a}
\lambda_{A_2, z} &= +2J_{z}\\
\label{eq:lEa}
\lambda_{E, FM} &= +2J_{x}\\
\label{eq:lEb}
\lambda_{E, AF} &= -J_{x}+\sqrt{3}D\\
\label{eq:lA}
\lambda_{A_1}=\lambda_{A_2, \perp} &= -J_{x}-\sqrt{3}D\\
\label{eq:lEc}
\lambda_{E, z} &= -J_{z}.
\end{align}
Except for the $E_{z}$ irrep, all other basis vectors correspond to normalized spins. Following the method detailed in section \ref{sec:howGS}, this allows for the direct determination of the ground-state for all parameters where $\lambda_{E, z}$ is not the minimal eigenvalue. The resulting phase diagram is given in Fig.~\ref{fig:PhD}. Portions of this phase diagram have been explored in the literature for classical Heisenberg spins, such as the Heisenberg antiferromagnet~\cite{Chalker1992,Zhitomirsky2008,Chern13b}, the XXZ model~\cite{Miyashita86a,Huse92a,Kuroda95a}, as well as Dzyaloshinskii-Moriya interactions~\cite{Elhajal2002}. In particular, this phase diagram has recently been shown to support a network of spin liquids with three-fold symmetry~\cite{Essafi16a}. This is why our goal in this section will be to present a comprehensive picture of the competing phases at play, in the context of the irreducible representations they are generated from.

%======================================================
%======================================================
\subsection{Long-Range Order}
\label{sec:XXZDMorder}

The various ordered phases presented below are categorized as a function of their global degeneracy and illustrated in Fig.~\ref{fig:PhD}. Their region of existence is easily calculated by ensuring that the corresponding eigenvalue(s) is (are) smaller than all the other ones [Eq.~(\ref{eq:lA2a}-\ref{eq:lEc})].

%======================================================
\subsubsection{$\mathbb{Z}_{2}$ degeneracy}
\label{sec:XXZDMA2a}

The only ground-state with $\mathbb{Z}_{2}$ degeneracy -- generated by 
time-reversal symmetry -- is the out-of-plane ferromagnetic phase, $A_{2z}$.

%======================================================
\subsubsection{O(2) degeneracy}

The XXZDM model is invariant by continuous global spin 
rotations around the $z$-axis. 
In other words, any ground-state with finite in-plane spin components 
possesses (at least) 
a global O(2) degeneracy. This is the case for the $E_{FM}$, $E_{AF}$ and $A=A_{1}\oplus A_{2\perp}$ phases, which are respectively stabilized by in-plane ferromagnetic, negative DM and positive DM interactions.

The O(2) degeneracy persists at the frontiers between the $E_{z}$ region and one of these ordered phases ($E_{FM}$, $E_{AF}$ or $A$), such as for example for $2 |D|/\sqrt{3}<J_{z}=-2 J_{x}$ [see the borders of the $E_{z}$ triangle in Fig.~\ref{fig:PhD}.$(c)$]. The reason why there is no enhancement of degeneracy at these frontiers is because $E_{z}$ order cannot co-exist with only one of the $E_{FM}$, $E_{AF}$ or $A$ phases. On the other hand, in presence of two other irreps, the co-existence with $E_{z}$ is possible; this corresponds to the Heisenberg antiferromagnet and equivalent models, which will be discussed in section \ref{sec:XXZDMSL}.

%======================================================
\subsubsection{O(3) degeneracy}

At the frontier between the $E_{FM}$ and $A_{2z}$ phases, $J_{x}=J_{z}<-|D|/\sqrt{3}$, the ground-state is ferromagnetic with O(3) degeneracy. Using the threefold symmetry in parameter space of the XXZDM model \cite{Essafi16a}, the same O(3) degeneracy holds for the three borders of the $A_{2z}$ triangle in Fig.~\ref{fig:PhD}.$(b)$, albeit with different spin configurations. These different spin configurations correspond to the umbrella states -- or variants thereof -- of the regular magnetic orders of Ref.~[\onlinecite{Messio2011}]. It is noticeable to find these regular magnetic orders in the highly anisotropic XXZ model with Dzyaloshinskii-Moriya considered here, thanks to an accidental O(3) degeneracy of the Hamiltonian.

All spin configurations are obtained from Eq.~(\ref{eq:Sop}), while imposing
\begin{eqnarray}
m_{A_{2z}}^{2}+\mathbf{m}_{I}^{2}=1 \quad {\rm with } \quad I\in\{A,E_{FM},E_{AF}\}.
\end{eqnarray}
At the $A_{2z}\oplus A$ and $A_{2z}\oplus E_{AF}$ frontiers, the out-of-plane ferromagnetism of the umbrella states conveys a finite scalar chirality
\begin{eqnarray}
\kappa=\mathbf{S}_{0} \cdot\left(\mathbf{S}_{1} \times \mathbf{S}_{2}\right)
\label{eq:scalarchiral}
\end{eqnarray}
to the ground-state manifold.\\

%======================
\renewcommand{\arraystretch}{2}
\begin{table*}
\begin{center}
\begin{tabular}{||C||C|C|C||}
\hhline{|t:=:t===:t|}
\textrm{Common features}
& \textrm{Positive DM}
& \textrm{Zero DM}
& \textrm{Negative DM}\\
\textrm{of the ground-states}
& 0<D=-\sqrt{3}\, J_{x}
& D=0 \quad\&\quad 0<J_{x}
& D=\sqrt{3}\, J_{x}<0\\
\hhline{|:=::=|=|=:|}
\textrm{Coulomb phase}
& \underline{E_{z},E_{FM},A} \quad [\mathcal{X}^{-}]
& \underline{E_{z},A,E_{AF}} \; [\textrm{HAF}]
& \underline{E_{z},E_{AF},E_{FM}} \quad [\mathcal{X}^{+}]\\
\multirow{2}{2.5cm}{\centering
$\mathcal{H}_{\Delta}=\dfrac{J_{z}}{2}(\mathbf{B}^{2}-3)$}
& 0<J_{z}=-2 J_{x}
& 0<J_{z}=J_{x}
& 0<J_{z}=-2 J_{x}\\
\multirow{2}{2.5cm}{\centering
The conserved flux $\mathbf{B}$ is given by:}
&\left(\begin{matrix}
-\frac{\sqrt{3}}{2} S_0^x - \frac{1}{2} S_0^y + \frac{\sqrt{3}}{2} S_1^x - \frac{1}{2} S_1^y + S_2^y \\
\frac{1}{2} S_0^x - \frac{\sqrt{3}}{2} S_0^y + \frac{1}{2} S_1^x + \frac{\sqrt{3}}{2} S_1^y - S_2^x \\
S_0^z + S_1^z + S_2^z
\end{matrix} \right)
&\left(\begin{matrix}
S_0^x + S_1^x + S_2^x \\
S_0^y + S_1^y + S_2^y \\
S_0^z + S_1^z + S_2^z
\end{matrix} \right)
&\left(\begin{matrix}
-\frac{\sqrt{3}}{2} S_0^x + \frac{1}{2} S_0^y + \frac{\sqrt{3}}{2} S_1^x + \frac{1}{2} S_1^y - S_2^y \\
\frac{1}{2} S_0^x + \frac{\sqrt{3}}{2} S_0^y + \frac{1}{2} S_1^x - \frac{\sqrt{3}}{2} S_1^y - S_2^x \\
S_0^z + S_1^z + S_2^z
\end{matrix} \right)
\\
\hhline{||----||}
\textrm{three-coloring}
& \underline{E_{FM},A} \quad [\textrm{XXZ}^{-}]
& \underline{A,E_{AF}} \; [\textrm{XXZ}]
& \underline{E_{AF},E_{FM}} \quad [\textrm{XXZ}^{+}]\\
\times\;\textrm{global O(2)}
& -|J_{x}| <J_{z}<2|J_{x}|
& -J_{x}/2 <J_{z}<J_{x}
& -|J_{x}| <J_{z}<2|J_{x}|\\
\hhline{||----||}
\textrm{three-coloring}
& \underline{E_{FM},A,A_{2z}} \quad [\textrm{FDM}^{-}]
& \underline{A,E_{AF},A_{2z}} \; [\textrm{XXZ}_{0}]
& \underline{E_{AF},E_{FM},A_{2z}} \quad [\textrm{FDM}^{+}]\\
\times\;\textrm{global O(3)}
& J_{x} = J_{z}<0
& -J_{x}/2 = J_{z}<0
& J_{x} = J_{z}<0\\
\hhline{|b:=:b===:b|}
\end{tabular}
\end{center}
\caption{
Summary of the network of extensively degenerate ground-states found in Ref.~[\onlinecite{Essafi16a}], presented in light of the irreps from which they are generated. These classical spin liquids sit at the frontier between ordered phases [section \ref{sec:XXZDMorder}]; their domain of existence at zero temperature is given in each case. There are three different branches in the network, with positive, negative and zero Dzyaloshinskii-Moriya (DM) couplings, corresponding to the three columns of the table. Going from one branch to the other is done by local transformations~\cite{Essafi16a}. These transformations can be rationalized here as a permutation of the bi-dimensional irreps with in-plane spin components: $E_{FM}$, $E_{AF}$ and $A$ [Eq.~(\ref{eq:EaEbA})]. Each pair of these irreps generates an extensive manifold of configurations that can be mapped onto the three-coloring problem, with additional O(2) global symmetry (third row). If the out-of-plane ferromagnetic irrep $A_{2z}$ is added, then the global degeneracy becomes O(3) (fourth row). If on the other hand the antiferromagnetic irrep $E_{z}$ is added, then one obtains a Coulomb phase defined by an emergent divergence-free field $\mathbf{B}$ (second row). The zero-temperature ground-state of the Heisenberg antiferromagnet (HAF) is described by the $E_{z}, A$ and $E_{AF}$ irreps ($0<J_{x}=J_{z}, D=0$). The names of the models are given in brackets, as defined in Ref.~[\onlinecite{Essafi16a}].}
\label{tab:SL}
\end{table*}
%======================

%======================================================

We should conclude this discussion with a few words about the finite-temperature physics. The Mermin-Wagner-Hohenberg theorem prevents any symmetry breaking phase transition at finite temperature in the Heisenberg ferromagnet (HFM) $(J_{x}=J_{z}<0$, $D=0$). By symmetry of our phase diagram, the models equivalent to the HFM with parameters
\begin{eqnarray}
J_{x}=-\dfrac{1}{2}J_{z}>0,\quad D=\pm\dfrac{\sqrt{3}}{2} J_{z}
\label{eq:equivHFM}
\end{eqnarray}
have the same energy excitations than the HFM, and are thus also protected by the Mermin-Wagner-Hohenberg theorem from ordering, despite their apparent anisotropy.

%======================================================
%======================================================
\subsection{XXZDM: classical spin liquids}
\label{sec:XXZDMSL}

%======================================================
\subsubsection{Three-fold mapping}
\label{ssec:3foldmap}

The phases discussed in section \ref{sec:XXZDMorder} are long-range ordered with wavevector $\mathbf{q}=0$. Once the orientation of one spin is known, the spin configuration of the entire lattice is fixed. This is not the case anymore when the ground-state is generated by combinations of $E_{FM}$, $E_{AF}$ and $A$ basis vectors. Such combinations give rise to a network of extensively degenerate phases \cite{Essafi16a}. This network is robust for both classical and quantum spins, and its branches are related to each other via a three-fold symmetry~\cite{Essafi16a} which is also valid for triangular lattices \cite{Momoi1992}. This mapping follows a similar motivation than for XXZ chains, where DM couplings can be ``erased'' by a local rotation and twisted boundary conditions \cite{Bocquet01a}. On kagome, twisted boundary conditions are not necessary because of specific choices of rotations. Here we shall clarify how these classical spin liquids can be understood from the point of view of their irreps, as summarised in Table \ref{tab:SL}.

%======================
\begin{figure}[t]
\begin{center}
\includegraphics[width=\columnwidth]{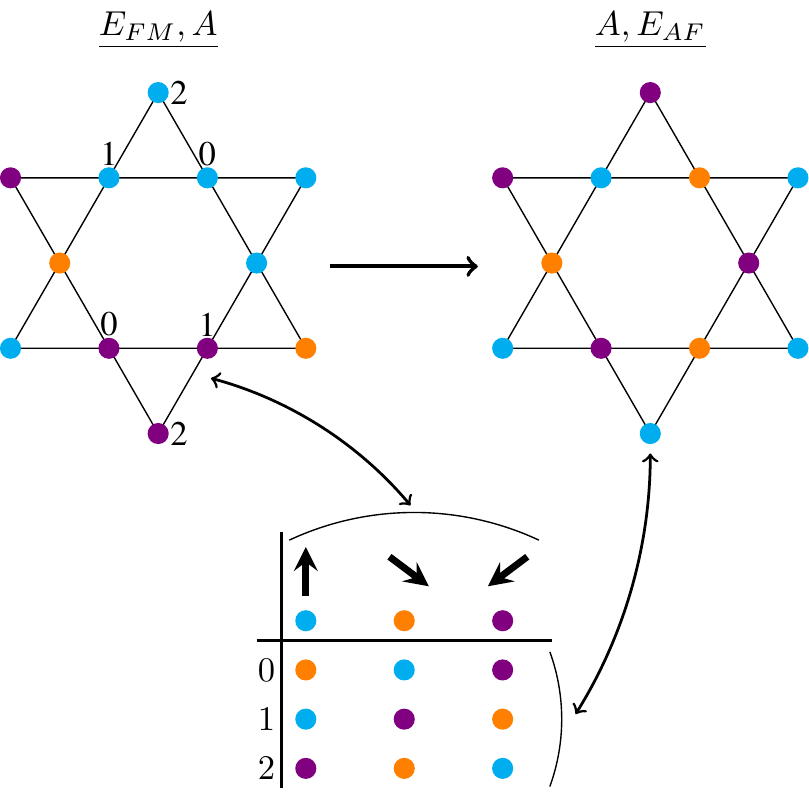}
\end{center}
\caption{All spin configurations generated by the $(A,E_{AF})$, $(E_{AF},E_{FM})$ or $(E_{FM},A)$ pair of irreps correspond to a three-color mapping on the kagome sites, or equivalently on the honeycomb bonds. While this coloring is straightforward in the former case~\cite{Huse1992} (top right), it is less so if the ferromagnetic $E_{FM}$ states are involved (top left). The mapping is given by the matrix at the bottom. For example for a spin on sublattice 1 (resp. 0 or 2) in a ferromagnetic triangle of violet color, the corresponding color is orange (resp. violet or cyan).
}
\label{fig:3ColourMapping}
\end{figure}
%======================

Let us consider the $(A,E_{AF})$ pair as a working example. These irreps generate the ground-state manifold of the antiferromagnetic XXZ model, which can be extended to ferromagnetic $J_{z}<0$ [Fig.~\ref{fig:PhD}]. It is well-known that this ground-state manifold can be mapped onto the three-coloring problem, whose mapping is unique up to a global O(2) rotation of the spins~\cite{Huse1992}. Indeed, all spins lie in the $xy$-plane and  make a 120$^{\circ}$ angle with their neighbours [Figs.~\ref{fig:PhD}.$(d)$ and \ref{fig:3ColourMapping}]. By symmetry, the same mapping also holds for the two other pairs of irreps
\begin{eqnarray}
(A,E_{AF}) \leftrightarrow (E_{AF},E_{FM}) \leftrightarrow (E_{FM},A).
\label{eq:EaEbA}
\end{eqnarray}
The correspondence between a configuration with ferromagnetic $E_{FM}$ states and the three-coloring model is given in Fig.~\ref{fig:3ColourMapping}.\\

The pairs of irreps in Eq.~(\ref{eq:EaEbA}) generates the ground-states on three lines of parameter space, 
which end in contact either with the $A_{2z}$ phase, or with the $E_{z}$ phase.

In the former case, the global O(2) degeneracy is enhanced to O(3). At zero temperature and in presence of Dzyaloshinskii-Moriya interactions, this symmetry enhancement confers a scalar chirality to the classical spin liquid \cite{Essafi16a}. In the irrep language, this corresponds to ground-states described by the $(E_{AF},E_{FM},A_{2z})$ or $(E_{FM},A,A_{2z})$ irreps [Table \ref{tab:SL}]. Remarkably, the classical degeneracy of these models has recently been shown to persist for quantum spins in every non-trivial $S^{z}$ sector \cite{Changlani17a}.

When the pair of irreps of Eq.~(\ref{eq:EaEbA}) are coupled to the $E_{z}$ phase, the zero-temperature ground-state manifold supports an emergent classical Coulomb phase, characterised by either antiferromagnetic or ferromagnetic pinch points in the structure factor~\cite{Essafi16a}. The trio of irreps, $(A,E_{AF},E_{z})$, corresponds to the emergent Coulomb phase of the canonical Heisenberg antiferromagnet (HAF).\\

%======================================================
\subsubsection{Along the XXZ line inside the $E_{z}$ region}
\label{ssec:XXZEc}

If one continues along the XXZ line towards the Ising antiferromagnet ($0<J_{x}<J_{z}$ and $D=0$), the minimal eigenvalue corresponds to the $E_{z}$ phase. Since the $E_{z}$ eigenstates are made of non-normalised spins, it means that irreps with higher eigenvalues are also populated. In increasing order, the excited eigenvalues are $E_{AF}\oplus A$ (degenerate), then $E_{FM}$ and finally $A_{2z}$ [Eqs.~(\ref{eq:lA2a}-\ref{eq:lEc})]. The ground-states in this region are known to bear a finite magnetisation \cite{Miyashita86a,Kuroda95a}, which can be
\begin{itemize}
\item either in plane due to the $E_{FM}$ component.\\
A typical ground state configuration is [\onlinecite{Miyashita86a}]\\ 
$\mathbf{S}_{0}=(1,0,0), \mathbf{S}_{1}=(-c,0,s), \mathbf{S}_{2}=(-c,0,-s)$\\
where $c=J_{x}/(J_{z}+J_{x})$ and $s=\sqrt{1-c^{2}}.$
\item or out of plane due to the $A_{2z}$ component.\\
A typical ground state configuration is [\onlinecite{Miyashita86a}]\\ 
$\mathbf{S}_{0}=(0,0,1), \mathbf{S}_{1}=(s',0,-c'), \mathbf{S}_{2}=(-s',0,-c')$\\
where $c'=J_{z}/(J_{z}+J_{x})$ and $s=\sqrt{1-c'^{2}}.$
\end{itemize}
This means that all eigenstates can be populated in the ground-state. In particular the $E_{AF}$ and $A$ order parameters, which correspond to the second lowest eigenvalue, always take a finite value. These are the same irreps responsible for the tricolouring spin liquid along the XXZ line for $J_{z}<J_{x}$, away from the Ising limit.

%======================
\begin{figure*}
\Large\hspace{1.1cm}$D=-\sqrt{3}/2$\hspace{1.7cm}$D=-0.4$\hspace{2.4cm}$D=0$\hspace{2.4cm}$D=\sqrt{3}/2$
\centering\includegraphics[width=18cm]{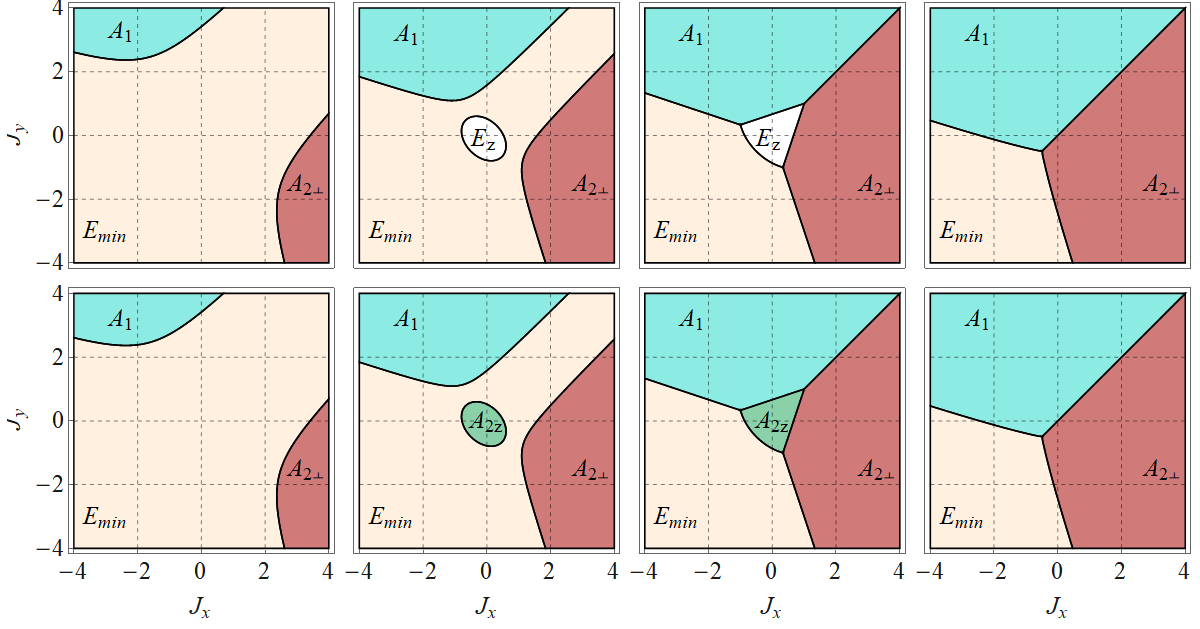}
\caption{Regions of the XYZDM parameter space $(J_{x},J_{y}$) where the lowest eigenvalue $\lambda_{I_{0}}$ corresponds to a given irrep $I_{0}\in\{A_{1}$ (blue), $A_{2\perp}$ (red), $A_{2 z}$ (green), $E_{min}$ (beige), $E_{z}$ (white)$\}$. The top and bottom panels correspond to $J_{z}=1$ and $J_{z}=-0.5$ respectively. The values of $D$ are given at the top of the figure. For these values of $J_{z}$, the $E_{z}$ and $A_{2z}$ regions appear as ground-states of the XYZDM model for $|D|<\sqrt{3}/2$. The $J_{x}=J_{y}$ line acts as a mirror in parameter space [section \ref{sec:invA1A2b}].}
\label{fig:XYZDMPD}
\end{figure*}
%======================

Here an interesting similarity appears with the quantum model. There are indeed strong indications \cite{Lauchli15a,He15a,He2015,Zhu15a,Hu15a,Lauchli16a} that the ground-state of the XXZ model  for quantum spins $S=1/2$ supports a quantum spin liquid, and that this quantum spin liquid remains in the same phase for the entire range of positive values of $J_{x}$ and $J_{z}$ ($D=0$). Our present work cannot explain this quantum phenomenon, but it brings a classical intuition. Along the XXZ line with $J_{x}, J_{z}>0$, the two lowest eigenvalues are always $\lambda_{E, z}$ and $\lambda_{E, AF}=\lambda_{A}$. They cross at the Heisenberg antiferromagnetic point ($J_{x}=J_{z}$), but the $E_{AF}\oplus A$ irreps are responsible for a (tricolouring) classical extensive degeneracy over the entire range of parameters. This is because they correspond to the lowest eigenvalue for $J_{z}<J_{x}$ (towards the XY limit), and they co-exist with the non-normalised $E_{z}$ irrep for $J_{x}<J_{z}$ (towards the Ising limit).

%======================================================
%======================================================
%======================================================
\section{The generic XYZDM model}
\label{sec:XYZDM}

In this section, the constraint (\ref{eq:Jxy}) is relaxed, giving rise to the XYZDM Hamiltonian (\ref{eq:JbasisB}), which has been diagonalised in Eq.~(\ref{eq:hamQ}). This model is described by four independent coupling parameters $\{J_{x},J_{y},J_{z},D\}$.

As opposed to the XXZDM model studied in section \ref{sec:XXZDM}, there is no in-plane O(2) invariance anymore. The main consequences of this broken symmetry are double.
\begin{enumerate}
\item The eigenvalues of the $A_{1}$ and $A_{2\perp}$ irreps are not degenerate [Eqs.~(\ref{eq:lambda-A1},\ref{eq:lambda-A2b})].
\item The $E_{FM}$ and $E_{AF}$ basis vectors are not eigenvectors of the coupling matrix $\mathcal{\hat J}$. The new eigenvectors correspond to $E_{\alpha}$ and $E_{\beta}$, as defined in Eq.~(\ref{eq:malpha-mbeta}), whose spins are not normalized. Furthermore, $E_{\alpha}$ and $E_{\beta}$ cannot be degenerate.
\end{enumerate}
Properties (1) and (2) are linked; it is not possible to have one without the other
\begin{eqnarray}
\lambda_{A_1}\neq \lambda_{A_2, \perp}
\;\Leftrightarrow\;
\lambda_{E, mix}\neq 0
\;\Leftrightarrow\;
\lambda_{E_{\alpha}} \neq \lambda_{E_{\beta}}.
\label{eq:linked}
\end{eqnarray}
For convenience let us define the eigenvalues
\begin{eqnarray}
%\begin{array}{ll}
&&\lambda_{E min}=\min[\lambda_{E,\alpha},\lambda_{E,\beta}] \nonumber  \\
&&=\frac{1}{2} \left( \lambda_{E, FM}+\lambda_{E, AF} -\sqrt{(\lambda_{E, FM}-\lambda_{E, AF})^2+\lambda_{E, mix}^2 } \right) \nonumber \\
&&\lambda_{E max}=\max[\lambda_{E,\alpha},\lambda_{E,\beta}] \nonumber \\
&&=\frac{1}{2} \left( \lambda_{E, FM}+\lambda_{E, AF} +\sqrt{(\lambda_{E, FM}-\lambda_{E, AF})^2+\lambda_{E, mix}^2 } \right) \nonumber \\
%\end{array}
\label{eq:minmax}
\end{eqnarray}
of the corresponding $E_{min}$ and $E_{max}$ irreps.\\

The additional ground-states permitted by the extra degree of freedom, $J_{x}\neq J_{y}$, comes from the newly possible combinations of irreps that were absent in the XXZDM model. 
One should be cautious though that not all combinations of irreps represent a possible ground-state of the XYZDM model. A given combination of irrep means degeneracy between their eigenvalues, which implies constraint(s) on the parameters $\{J_{x},J_{y},J_{z},D\}$. Within this constrained parameter region, one needs to check if the degenerate combination of irreps possesses the lowest eigenvalue $\lambda$ [Eqs.~(\ref{eq:lambda-A1},\ref{eq:lambda-A2a},\ref{eq:lambda-A2b},\ref{eq:lambda-Ec},\ref{eq:lambda-Ealpha},\ref{eq:lambda-Ebeta})]. As illustrated in Fig.~\ref{fig:XYZDMPD}, and discussed in detail in the present section \ref{sec:XYZDM}, this is true for a broad diversity of unexplored phases.

%======================================================
%======================================================
\subsection{Symmetry between $A_{1}$ and $A_{2\perp}$}
\label{sec:invA1A2b}

In the three-dimensional parameter space $\left(\dfrac{J_{x}}{D},\dfrac{J_{y}}{D},\dfrac{J_{z}}{D}\right)$, the two-dimensional subspace defined by $J_{x}=J_{y}$ is a mirror symmetry in the thermodynamic properties of the XYZDM model. Indeed, Eqs.~(\ref{eq:lambda-A1}-\ref{eq:lambda-Ec}) and (\ref{eq:phi}) respect the following invariance
\begin{eqnarray}
\begin{array}{ll}
A_{1}\leftrightarrow A_{2\perp}\\
J_{x}\leftrightarrow J_{y}\\
\phi \leftrightarrow -\phi
\end{array}
\label{eq:invA1A2b}
\end{eqnarray}
It means that all results obtained for the $A_{1}$ irrep are directly applicable to $A_{2\perp}$, and vice-versa.

%======================================================
%======================================================
\subsection{Intrinsic chiral asymmetry}
\label{sec:chiasym}

The vector chirality for a triangle is given by~\cite{Villain77b,Kawamura84a}
\begin{eqnarray}
{\boldsymbol\kappa}=\dfrac{2}{3\sqrt{3}}\left(\mathbf{S}_{0} \times \mathbf{S}_{1}+\mathbf{S}_{1} \times \mathbf{S}_{2}+\mathbf{S}_{2} \times \mathbf{S}_{0}\right).
\label{eq:chivec}
\end{eqnarray}
The $z$ component of the vector chirality, $\kappa_{z}$, is the conjugate variable of the Dzyaloshinskii-Moriya parameter $D$ of Eq.~(\ref{eq:DM}). Hence, $\kappa_{z}$ takes a saturated value for the ground-states induced by Dzyaloshinskii-Moriya interactions~\cite{Elhajal2002}, namely the $A_{1}, A_{2\perp}$ ($\kappa_{z}=- 1$) and $E_{AF}$  ($\kappa_{z}=+ 1$) states [Fig.~\ref{fig:irrep}].

The XXZDM model ($J_{x}= J_{y}$) is symmetric under sign reversal of $D$; it means that the contributions of the $E_{AF}$ and $A=A_{1}\oplus A_{2\perp}$ are exchanged in the spin configurations of Eq.~(\ref{eq:Sop}) when $D\rightarrow -D$.

In the XYZDM model on the other hand ($J_{x}\neq J_{y}$), the DM-reversal symmetry is broken. For large positive $D$, the $A_{1}$ and $A_{2\perp}$ states are respectively favoured by positive $J_{x}$ and $J_{y}$, with global $\mathbb{Z}_{2}$ degeneracy. This means that the two basis vectors with negative chirality ($\kappa_{z}=- 1$) can be differentiated energetically. On the other hand, the $E_{AF}$ irrep with positive chirality ($\kappa_{z}=+ 1$) becomes mixed with the non-chiral $E_{FM}$ irrep into states which are not normalised anymore [Eq.~(\ref{eq:malpha-mbeta})]. The fate of the XYZDM model is thus particularly asymmetric between positive and negative values of DM interactions.\\

This intrinsic asymmetry between positive and negative chirality comes from the fact that any transformation of the $C_{3v}$ group given in Eq.~(\ref{eq:D3}) is at the same time a permutation of the sites within a triangle, and a rotation of the spin orientations. Let us consider Fig.~\ref{fig:irrep}.
\begin{itemize}
\item For the $A_{1}$ and $A_{2\perp}$ states, a clockwise permutation of the sites comes with a clockwise rotation of the spin orientations. Any state with negative chirality $\kappa_{z}=- 1$ is left invariant under a $C_{3}$ transformation, and a uni-dimensional irrep is sufficient to ensure invariance.
\item For the $E_{AF}$ states, a clockwise permutation of the sites comes with a \textit{counterclockwise} rotation of the spin orientations; any state with positive chirality $\kappa_{z}=+ 1$ is modified under a $C_{3}$ transformation, and a two-dimensional subspace becomes necessary to recover invariance.
\end{itemize}
This chiral asymmetry is not unique to kagome though. The physics of direct ($D>0$) and indirect ($D<0$) Dzyaloshinskii-Moriya interactions on the pyrochlore antiferromagnet are known to be qualitatively different, both at zero and finite temperatures~\cite{Elhajal05a,Canals2008,Chern10b,Yan17a} [Fig.~\ref{fig:pyro}]. In analogy with kagome, direct DM interactions on pyrochlore are known to favour the all-in all-out ordered phase, which transforms according to the $A_{2}$ uni-dimensional irrep. As for indirect DM interactions, they favour the so-called $\Gamma_{5}$ configurations, which transforms according to the $E$ two-dimensional irrep.

A specificity of the kagome lattice is actually that this chiral asymmetry disappears for a large portion of coupling parameters, namely the XXZDM model when $J_{x}=J_{y}$.

%======================================================
%======================================================
\subsection{Long-range orders with only trivial time-reversal symmetry}
\label{sec:Z2}

%======================================================
\subsubsection{Ferromagnetism: $A_{2z} \oplus E_{min}$}
\label{sec:EaorEbA2a}

The $A_{2z}$ states are incompatible with the $E_{\alpha}$ or $E_{\beta}$ states. This is because the ferromagnetic $A_{2z}$ contribution provides the same $S_{i=0,1,2}^{z}=S^{z}$ component to the three spins in the triangle. Since spin normalization imposes
\begin{eqnarray}
|\mathbf{S}_{i}^{\perp}|=\sqrt{1-(S^{z})^{2}},\quad \forall i=0,1,2\quad,
\end{eqnarray}
the three in-plane spin components $\mathbf{S}_{i=1,3}^{\perp}$ have to be of the same norm. This is not possible for the non-normalized $E_{min}$ basis vectors [section \ref{sec:diagham}]. The only solution is $S^{z}=\pm 1$. At the frontiers in the phase diagram where $\lambda_{A_{2z}}=\lambda_{E_{min}}$, out-of-plane ferromagnetism ($A_{2z}$ order) is energetically favoured.

%======================================================
\subsubsection{Vector chirality: $A_{1}$ or $A_{2\perp}$ ($\oplus E_{z}$)}
\label{sec:A1orA2b}

In addition to the out-of-plane ferromagnetic $A_{2z}$ order, new ordered phases with global $\mathbb{Z}_{2}$ degeneracy appear in the XYZDM model. They correspond to either $A_{1}$ or $A_{2\perp}$ order and carry a saturated vector chirality ${\boldsymbol\kappa}$. These phases are ground-states of a large portion of the phase diagram, as illustrated in Fig.~\ref{fig:XYZDMPD}.

Furthermore, the $A_{1}$ or $A_{2\perp}$ states are incompatible with the $E_{z}$ states. If a spin configuration is a linear combination of the $E_{z}$ and either the $A_{1}$ or the  $A_{2\perp}$ basis vectors [Eq.~(\ref{eq:Sop})], then imposing normalisation of all spins makes the $E_z$ contribution null. It means that the $A_{1}$ or $A_{2\perp}$ orders persist up to, and including, the frontiers with $E_z$.

%======================================================
%======================================================
\subsection{Scalar-chiral order with global O(2) invariance:\\ $(A_{1}$ or $A_{2\perp}) \oplus A_{2z}$}
\label{sec:A1orA2bA2a}

At the frontiers between the ferromagnetic $A_{2z}$ states and one of the other uni-dimensional irreps, $A_{1}$ or $A_{2\perp}$, the ground-states are obtained by erasing all the other order parameters, resulting in
\begin{align}
m_{A_{2z}}^2 + m_{I}^2 = 1,
\end{align}
with $I=\{A_{1},A_{2\perp}\}$. Such ground-state manifold has a O(2) degeneracy, parametrized by $\upsilon$
\begin{align}
\left\{\begin{array}{ll}
m_{A_{2z}} &= \cos \upsilon \\
m_{I} &= \sin \upsilon
\end{array}
\right. .
\end{align}
Injecting these solutions into Eq.(\ref{eq:Sop}) leads to the following normalized spin configurations with long-range $\mathbf{q}=0$ order and finite scalar chirality
\begin{align}
A_{1}\oplus A_{2z}&
\left\{\begin{array}{ll}
{\bf S}_0 &= \left( \frac{1}{2} \sin \upsilon, \frac{\sqrt{3}}{2} \sin \upsilon, \cos \upsilon \right) \\
{\bf S}_1 &= \left( \frac{1}{2} \sin \upsilon, -\frac{\sqrt{3}}{2} \sin \upsilon, \cos \upsilon \right) \\
{\bf S}_2 &= \left( -\sin \upsilon, 0, \cos \upsilon \right)
\end{array}
\right.\\
\label{eq:A2perpA2z}
A_{2\perp}\oplus A_{2z}&
\left\{\begin{array}{ll}
{\bf S}_0 &= \left(-\frac{\sqrt{3}}{2} \sin \upsilon, \frac{1}{2}\sin \upsilon, \cos \upsilon \right) \\
{\bf S}_1 &= \left(\frac{\sqrt{3}}{2} \sin \upsilon, \frac{1}{2} \sin \upsilon, \cos \upsilon \right) \\
{\bf S}_2 &= \left(0, -\sin \upsilon, \cos \upsilon \right)
\end{array}\right.
\end{align}

%======================================================
%======================================================
\subsection{Stripe order with local $\mathbb{Z}_{8}$ degeneracy:\\$(A_{1}$ or $A_{2\perp}) \oplus E_{min}$}
\label{sec:Z8}

The $E_{min}$ irrep corresponds to non-normalized spin configurations [Eqs.~(\ref{eq:malpha-mbeta}), (\ref{eq:minmax})]. However, when combined with another irrep, it is {\it a priori} possible for a linear combinations of the two to respect the condition $|\mathbf{S}_{i}|^{2}=1$ for all spins $i$. The goal of this section is to prove this possibility for the $A_{1}$ and $A_{2\perp}$ irreps. While we will use $A_{2\perp}$ as an example, all arguments also directly apply to $A_{1}$ [section \ref{sec:invA1A2b}]. In section \ref{ssec:A1A2bEminA2a}, we will briefly discuss what happens at the frontier with out-of-plane ferromagnetism ($A_{2z}$ irrep).

%======================================================
\subsubsection{Spin configurations}
\label{ssec:Z8}
In this section, we consider Hamiltonians where the ground-states are linear combinations of the $A_{2\perp}$ and $E_{min}$ spin configurations, \textit{i.e.}
\begin{eqnarray}
&\lambda_{A_{2\perp}}=\lambda_{E min} < \lambda_{I\in\{A_{1},A_{2z},E_{z},E_{max}\}}
\label{eq:constraintA1Emin}\\
&\Rightarrow \mathbf{m}_{I\in\{A_{1},A_{2z},E_{z},E_{max}\}}=0.
\label{eq:zerom}
\end{eqnarray}
According to Eq.(\ref{eq:malpha-mbeta}),
\begin{eqnarray}
\label{eq:eta}
&\mathbf{m}_{E, max}=0 \Rightarrow {\bf m}_{E, AF} = \eta\; {\bf m}_{E, FM},\\
&\mathrm{with}\quad \eta =\left\{\begin{array}{ll}
 -\tan \phi\qquad &\mathrm{ if } \; E_{min}=E_\alpha\\
+ \cot \phi \qquad &\mathrm{ if }\; E_{min}=E_\beta
\end{array}
\right.
\nonumber
\end{eqnarray}
where $\eta$ is function of $\phi$ and thus depends on the coupling parameters $\{J_x,J_y,J_z,D\}$ of the Hamiltonian. In practice, the values of these parameters are constrained by Eq.~(\ref{eq:constraintA1Emin}). But since $\phi\in[-\frac{\pi}{4}:\frac{\pi}{4}]$ [Eq.~(\ref{eq:phi})], $\eta$ can {\it a priori} take any real values. This is why we will first consider the general case, $-\infty<\eta<\infty$. Then we will analyse the range of possible ground-states as a function of $\eta$, and calculate what are the corresponding parameters $\{J_x,J_y,J_z,D\}$ that respect the condition~(\ref{eq:constraintA1Emin}).\\

%======================
\begin{figure}[b]
\centering
\includegraphics[width=8.5cm]{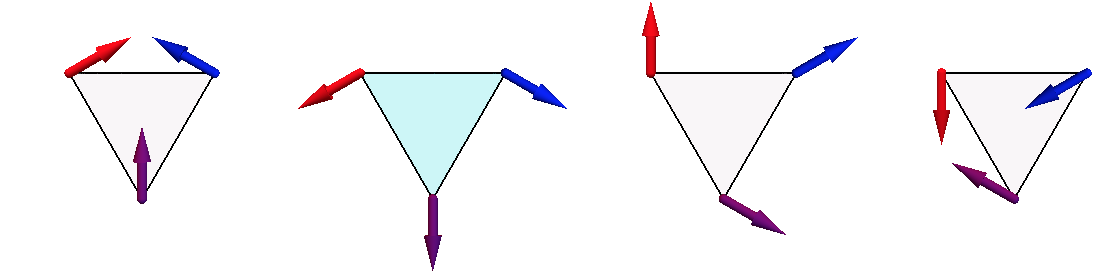}\\
\includegraphics[width=8.5cm]{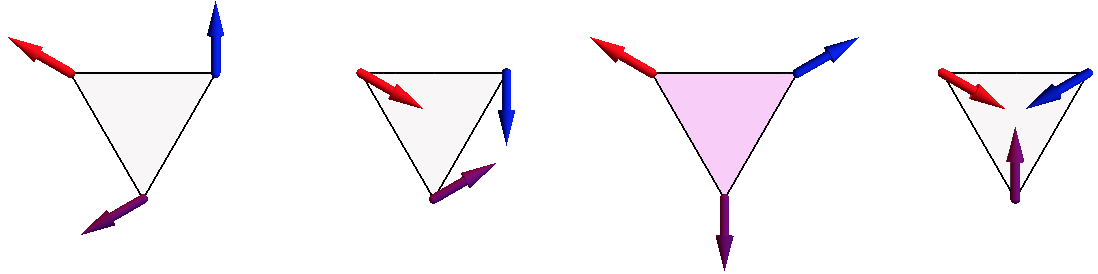}\\
\caption{Ground states for an arbitrary value of coupling parameters sitting at the frontier between $A_{2\perp}$ and $E_{min}$; $0<J_x=-\dfrac{J_y}{3}, \quad D=0, \quad J_y/2 < J_z < -J_y$ which corresponds to $\eta=-0.5$. See Eqs.~(\ref{eq:ups1}-\ref{eq:ups4}) for the spin configurations. The two states on the bottom right correspond to $A_{2\perp}$. Each of the $A_{2\perp}$ states can be ``paired'' with three other states, having one spin in common with the same orientation; see for example the violet spin of the two colored triangles. On the other hand, none of the six other states can be paired together. Hence, the $\mathbb{Z}_{8}$ degeneracy is divisible between two exclusive groups of four states, centred around each of the $A_{2\perp}$ states.}
\label{fig:Z8}
\end{figure}
%======================

Using Eqs.(\ref{eq:zerom}) and (\ref{eq:eta}), we are left with three variables $m_{A_{2\perp}}$, $m_{E,FM}^x$ and $m_{E,FM}^y$ which completely determine the spin configurations, as given in Eqs.~(\ref{eq:S0}-\ref{eq:S2}). Imposing the unit-length constraint gives a set of three non-linear equations with three unknown variables. Solving this set of equations gives the values of $m_{A_{2\perp}}$, $m_{E,FM}^x$ and $m_{E,FM}^y$, and thus the ensemble of ground-states. Since $m_{A_{2z}}=0=|\mathbf{m}_{E_z}|$, we know that $S_i^z=0$, and can restrict the spin configurations to in-plane components
\begin{eqnarray}
\Upsilon=\{S_0^x,S_0^y,S_1^x,S_1^y,S_2^x,S_2^y\}.
\end{eqnarray}
At the level of a triangle, the ground-state between $A_{2\perp}$ and $E_{min}$ is 8-fold degenerate
\begin{widetext}
\begin{align}
\label{eq:ups1}
\Upsilon^\pm_1=\pm(&
-\frac{\sqrt{3}\;\eta\;(2+\eta)}{2(1+\eta+\eta^2)},
\frac{2+2\eta-\eta^2}{2(1+\eta+\eta^2)},&
+\dfrac{\sqrt{3}\;\eta\;(2+\eta)}{2(1+\eta+\eta^2)},
\dfrac{2+2\eta-\eta^2}{2(1+\eta+\eta^2)},&
\qquad 0,
1&)\\
\label{eq:ups2}
\Upsilon^\pm_2=\pm(&
+\frac{\sqrt{3}\;(1+2\eta)}{2(1+\eta+\eta^2)},
\frac{1-2\eta-2\eta^2}{2(1+\eta+\eta^2)},&
\dfrac{\sqrt{3}}{2},
\dfrac{1}{2},&
+\dfrac{\sqrt{3}\;(1-\eta^2)}{2(1+\eta+\eta^2)},
\dfrac{1+4\eta+\eta^2}{2(1+\eta+\eta^2)}&)\\
\label{eq:ups3}
\Upsilon^\pm_3=\pm(&
-\dfrac{\sqrt{3}}{2},
\dfrac{1}{2},&
-\frac{\sqrt{3}\;(1+2\eta)}{2(1+\eta+\eta^2)},
\frac{1-2\eta-2\eta^2}{2(1+\eta+\eta^2)},&
-\dfrac{\sqrt{3}\;(1-\eta^2)}{2(1+\eta+\eta^2)},
\dfrac{1+4\eta+\eta^2}{2(1+\eta+\eta^2)}&)\\
\label{eq:ups4}
\Upsilon^\pm_4=\pm(&
-\dfrac{\sqrt{3}}{2},
\dfrac{1}{2},&
\dfrac{\sqrt{3}}{2},
\dfrac{1}{2},&
\qquad 0,
-1&)
\end{align}
\end{widetext}
Equivalently, the ground-state degeneracy is also 8-fold at the frontier between $A_{1}$ and $E_{min}$
\begin{widetext}
\begin{align}
\label{eq:ups1p}
\Upsilon'^\pm_1=\pm(&
\frac{-2+2\eta+\eta^2}{2(1-\eta+\eta^2)},
\frac{-\sqrt{3}\;\eta\;(-2+\eta)}{2(1-\eta+\eta^2)},&
\dfrac{-2+2\eta+\eta^2}{2(1-\eta+\eta^2)},
\dfrac{\sqrt{3}\;\eta\;(-2+\eta)}{2(1-\eta+\eta^2)},&
\qquad -1,
0&)\\
\label{eq:ups2p}
\Upsilon'^\pm_2=\pm(&
\frac{-(1+2\eta-2\eta^2)}{2(1-\eta+\eta^2)},
\frac{\sqrt{3}\;(1-2\eta)}{2(1-\eta+\eta^2)},&
-\dfrac{1}{2},
\dfrac{\sqrt{3}}{2},&
\dfrac{-(1-4\eta+\eta^2)}{2(1-\eta+\eta^2)},
\dfrac{\sqrt{3}\;(1-\eta^2)}{2(1-\eta+\eta^2)}&)\\
\label{eq:ups3p}
\Upsilon'^\pm_3=\pm(&
-\dfrac{1}{2},
-\dfrac{\sqrt{3}}{2},&
\frac{-(1+2\eta-2\eta^2)}{2(1-\eta+\eta^2)},
\frac{\sqrt{3}\;(-1+2\eta)}{2(1-\eta+\eta^2)},&
\dfrac{-(1-4\eta+\eta^2)}{2(1-\eta+\eta^2)},
\dfrac{-\sqrt{3}\;(1-\eta^2)}{2(1-\eta+\eta^2)}&)\\
\label{eq:ups4p}
\Upsilon'^\pm_4=\pm(&
\dfrac{1}{2},
\dfrac{\sqrt{3}}{2},&
\dfrac{1}{2},
-\dfrac{\sqrt{3}}{2},&
\qquad -1,
0&)
\end{align}
\end{widetext}

This is by itself a noticeable result. Indeed, we have here an extended region of parameters, at the frontier between the $E_{min}$ and $A_{2\perp}$ (or equiv. $A_{1}$) irreps [Fig.~\ref{fig:XYZDMPD}], with a local \textit{eight-fold} degeneracy. This discreteness is neither due to single-ion anisotropy, nor a symmetry-breaking magnetic field, nor the quantization of spins. It emerges naturally from a time-reversal invariant Hamiltonian with classical O(3) spins. For such models, a two-fold degeneracy is commonly induced by time-reversal symmetry. The degeneracy can be enhanced by the lattice symmetry: for example four-fold or six-fold for square or cubic lattices respectively. On kagome, the natural expectation would have been $\mathbb{Z}_{6}=\mathbb{Z}_3\otimes\mathbb{Z}_2$. And for higher symmetry ground-states, linear combinations of multiple classical orders usually allow for a continuous degree of freedom connecting the various ordered phases.

This is not the case here. The reason comes from the non-normalized irrep $E_{min}\in\{E_{\alpha},E_{\beta}\}$ which (i) prevents the continuous connection between multiple orders, on the basis that some of the orders are not physical while (ii) nonetheless allowing for a discrete number of physical linear combinations, \textit{i.e.} with normalized spins. These additional states respect the ``natural'' $\mathbb{Z}_{6}$ kagome symmetry. Once added to the pre-existing $A_{1}$ states, we get the $\mathbb{Z}_{8}$ degeneracy.

%======================================================
\subsubsection{Stripe order}
\label{ssec:stripes}
 
%======================
\begin{figure}[t]
\centering\includegraphics[width=6.5cm]{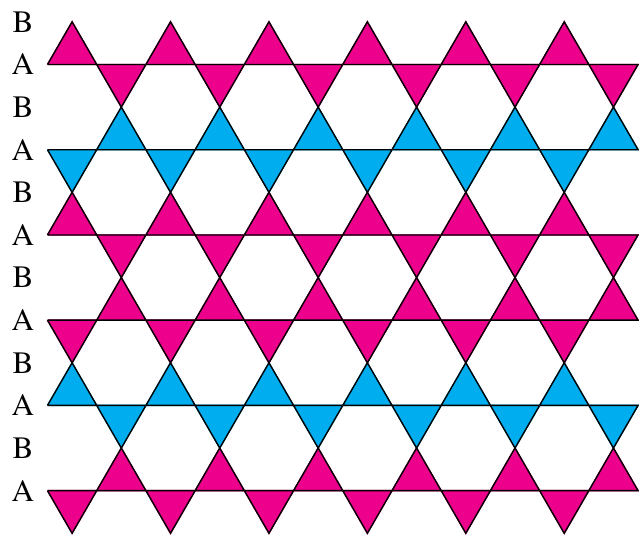}\\
\caption{Example of a stripe order emerging on the frontier between $E_{min}$ and $A_{2\perp}$ irreps. The magenta and cyan triangles correspond, for example, to the two coloured states of Fig.~\ref{fig:Z8}. Spins on the ``B'' lines are long-range ordered all over the lattice. As for spins on the ``A'' lines, they are long-range ordered in one direction (horizontal) but disordered in the other direction (vertical) where they can randomly take one out of two possible orientations. By symmetry, the stripes can also be diagonal.
}
\label{fig:stripe}
\end{figure}
%======================

Fig.~\ref{fig:Z8} provides a visual representation of the $\mathbb{Z}_{8}$ ground-states for an arbitrary value of coupling parameters at the frontier between $A_{2\perp}$ and $E_{min}$. Even if the ground-state is eight-fold degenerate, each sublattice (red, blue or violet spins) only has six possible spin orientations. Each of the spin orientations of a $A_{2\perp}$ state is also present in one of the other six states, creating pairs of states [Fig.~\ref{fig:Z8}]. The consequence of this pairing is a sub-extensive ground-state entropy, as explained below.

Imagine a horizontal line of $A_{2\perp}$ states on the kagome lattice, such as the bottom line of magenta triangles of Fig.~\ref{fig:stripe}. The pairing allows for the line of triangles just above to be one of two kinds: either the same $A_{2\perp}$ state, or the paired state sharing the same spin; see \textit{e.g.} the two coloured triangles in Fig.~\ref{fig:Z8}. Repeating the procedure gives rise to a stripe order, where each stripe can be of arbitrary width [Fig.~\ref{fig:stripe}]. By choosing another pair of states in Fig.~\ref{fig:Z8}, the stripes can be made diagonal. It is not possible to terminate a stripe in the bulk, because the non-$A_{2\perp}$ states are paired with one, and only one, other state. The resulting degeneracy of this ground-state is $\sim 2^{L}$ for a system of open boundaries and linear size $L$.\\

Since the ground-state configurations are not linked by an exact symmetry of the Hamiltonian, thermal fluctuations may lift the degeneracy between them at finite temperature, via an order-by-disorder mechanism.

%======================
\begin{figure}[t]
\centering\includegraphics[width=6.5cm]{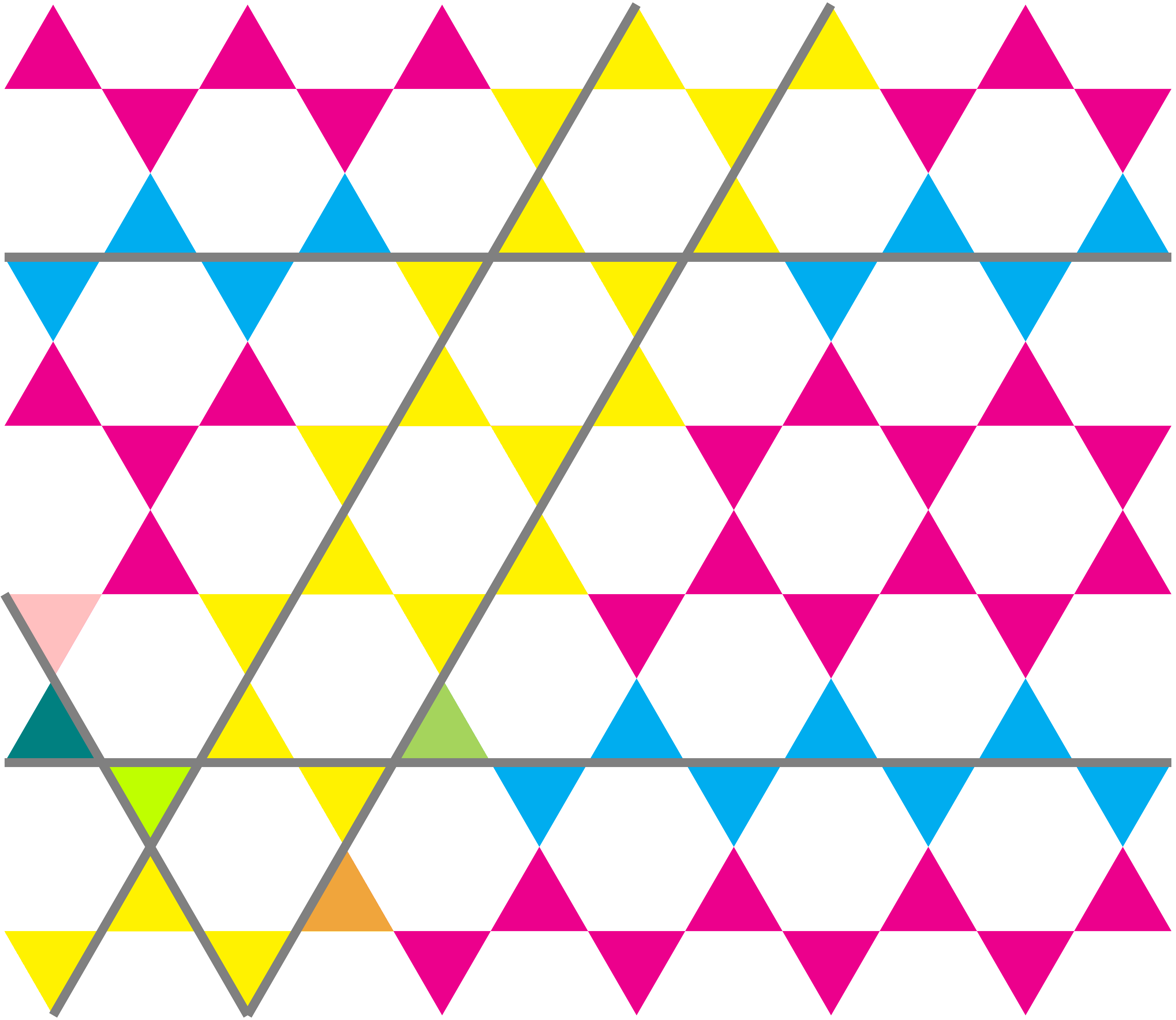}\\
\caption{Example of a configuration with crossing stripes, a possible ground-state for the parameters given in Eqs.(\ref{eq:Z8J1}) and (\ref{eq:Z8J2}). Each color corresponds to one of the eight degenerate ground-states of Fig.~\ref{fig:Z8star}. The grey lines are a guide to the eye for the position of the stripes. For each of the eight colours, the three edges of the triangle are covered by stripes in a different way.
}
\label{fig:cstripe}
\end{figure}
%======================

%======================================================
\subsubsection{Crossing stripes for high-symmetry Hamiltonians}
\label{ssec:cstripe}
 
Along the frontier between $E_{min}$ and ($A_{1}$ or $A_{2\perp}$), $\eta$ varies continuously [Eq.~(\ref{eq:eta})], allowing for a smooth deformation of the ground-state configurations given in Eqs.~(\ref{eq:ups1}-\ref{eq:ups4}) and (\ref{eq:ups1p}-\ref{eq:ups4p}). For example in the cyan triangle of Fig.~\ref{fig:Z8}, the orientation of the red and blue spins rotates in the kagome plane when varying $\eta$, while the violet spin remains fixed. It means that for specific values of $\eta$, these rotating spins can overlap with each other.
 This overlap provides more possibilities to connect the triangles next to each other on the kagome lattice, and thus a higher entropy. There are four specific values of $\eta$ with such high symmetry, two at the frontier with $A_{1}$, and another two at the frontier with $A_{2\perp}$.
\begin{eqnarray}
\eta^{\ast}=\pm\left(2-\sqrt{3}\right)^{\pm 1}
\label{eq:etastar}
\end{eqnarray}
The opposite values of $\eta$ comes from the symmetry between $A_{1}$ and $A_{2\perp}$ [section \ref{sec:invA1A2b}]. When combined with the symmetry between $E_{\alpha}$ and $E_{\beta}$ [Eq.~(\ref{eq:eta})], one gets the reciprocal values of $\eta^{\ast}$ in Eq.~(\ref{eq:etastar}). A word of caution, though. Considering the spin configurations of Eqs.~(\ref{eq:ups1}-\ref{eq:ups4}) and (\ref{eq:ups1p}-\ref{eq:ups4p}), one can find other ensemble of states of high symmetry, corresponding to other values of $\eta$. The corresponding values of $\{J_{x},J_{y},J_{z},D\}$, however, do not satisfy the constraint of Eq.~(\ref{eq:constraintA1Emin}) and the spin configurations do not correspond to ground-state configurations.

%======================
\begin{figure*}[t]
\centering
{\large  $J_x=-\dfrac{2-\sqrt{3}}{2+\sqrt{3}}J_y \qquad D=-(2-\sqrt{3})J_y \qquad -\dfrac{2 |J_y|}{2+\sqrt{3}} < J_z < \dfrac{4|J_y|}{2+\sqrt{3}}$\\
\vspace{0.2cm}
\tikz{\draw[dashed] (0,0)--(14,0);}\\\vspace{0.2cm}
a) Ground states at the frontier between $A_{2\perp}$ and $E_{\alpha}$ \quad $(J_{y}<0 \Rightarrow \eta=\sqrt{3}-2)$
\includegraphics[width=17cm]{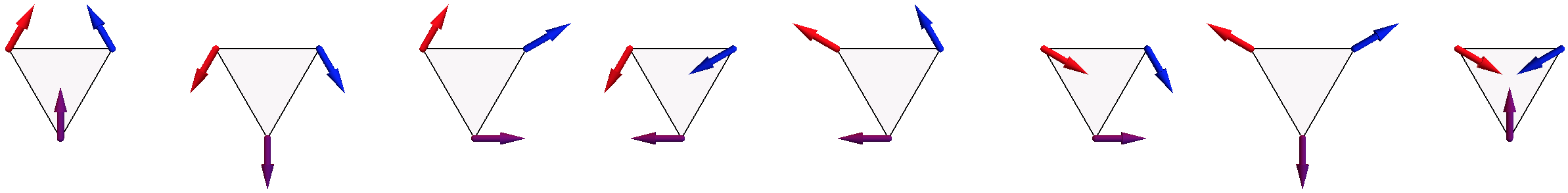}\\
\tikz{\draw[dashed] (0,0)--(14,0);}\\\vspace{0.2cm}
b) Ground states at the frontier between $A_{1}$ and $E_{\beta}$ \quad $(J_{y}>0 \Rightarrow \eta=\dfrac{1}{2-\sqrt{3}})$}
\includegraphics[width=17cm]{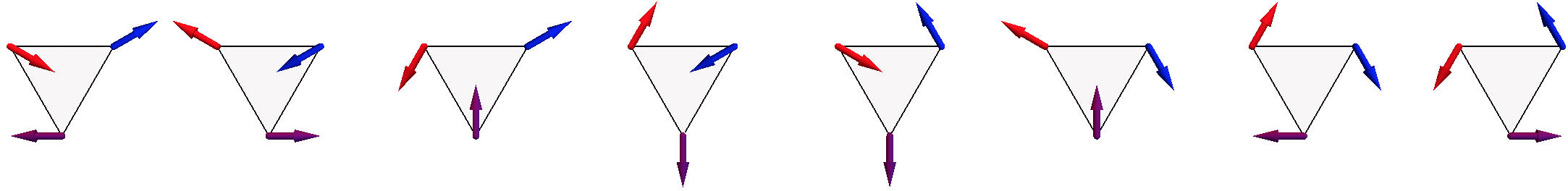}
\tikz{\draw[very thick] (0,0)--(18,0);}\\\vspace{0.2cm}
\vspace{0.2cm}
{\large  $J_y=-\dfrac{2-\sqrt{3}}{2+\sqrt{3}}J_x \qquad D=-(2-\sqrt{3})J_x \qquad -\dfrac{2 |J_x|}{2+\sqrt{3}} < J_z < \dfrac{4|J_x|}{2+\sqrt{3}}$\\
\vspace{0.2cm}
\tikz{\draw[dashed] (0,0)--(14,0);}\\\vspace{0.2cm}
c) Ground states at the frontier between $A_{2\perp}$ and $E_{\beta}$ \quad $(J_{x}>0\Rightarrow \eta=\dfrac{1}{\sqrt{3}-2})$
\includegraphics[width=17cm]{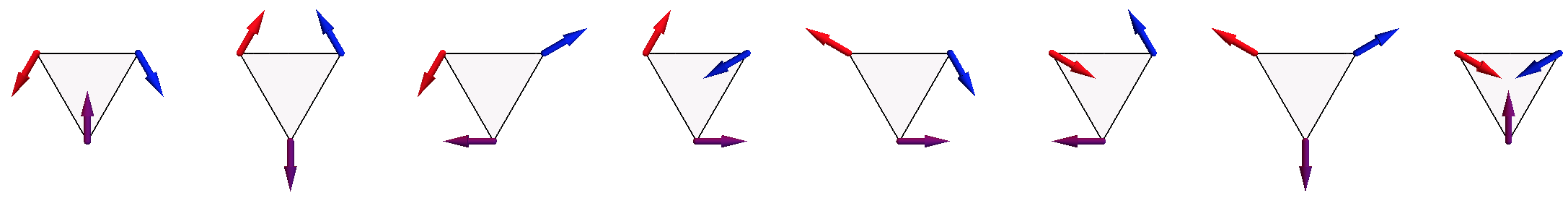}\\
\tikz{\draw[dashed] (0,0)--(14,0);}\\\vspace{0.2cm}
d) Ground states at the frontier between $A_{1}$ and $E_{\alpha}$ \quad $(J_{x}<0\Rightarrow\eta=2-\sqrt{3})$}
\includegraphics[width=17cm]{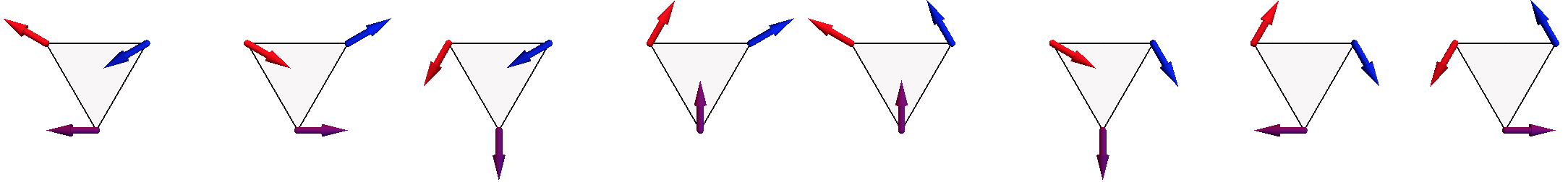}
\caption{In the XYZDM model, when the $E_{min}\in\{E_{\alpha},E_{\beta}\}$ irrep meets one of the one-dimensional antiferromagnetic irreps, $A_{1}$ or $A_{2\perp}$, there is a local $\mathbb{Z}_{8}$ degeneracy for each triangle [Fig.~\ref{fig:Z8}]. For special values of parameters on this frontier, there is an enhancement of the symmetry, where every ground-state shares a common spin orientation with three other ground-states. The four different sets of high-symmetry ground-states are displayed here, together with their parameter region. Injecting the corresponding value of $\eta$ in Eqs.~(\ref{eq:ups1}-\ref{eq:ups4}) and (\ref{eq:ups1p}-\ref{eq:ups4p}) gives the expression of the spin configurations. The values of $\eta$ are uniquely determined by the coupling parameters $\{J_{x},J_{y},J_{z},D\}$ [Eqs.(\ref{eq:phi}) and (\ref{eq:eta})]. The two states on the right correspond to either $A_{1}$ or $A_{2\perp}$.}
\label{fig:Z8star}
\end{figure*}
%======================

The spin configurations for the different values of $\eta^{\ast}$ are given in Fig.~\ref{fig:Z8star}. These configurations are ground-states of the XYZDM model for the following range of parameters

\begin{eqnarray}
\label{eq:Z8J1}\left\{
\begin{array}{ll}
J_x=-\dfrac{2-\sqrt{3}}{2+\sqrt{3}}J_y\\
D=-(2-\sqrt{3})J_y\\
-\dfrac{2 |J_y|}{2+\sqrt{3}} < J_z < \dfrac{4|J_y|}{2+\sqrt{3}}
\end{array}\right.
\end{eqnarray}

\begin{eqnarray}
\label{eq:Z8J2}\left\{
\begin{array}{ll}
J_y=-\dfrac{2-\sqrt{3}}{2+\sqrt{3}}J_x\\
D=-(2-\sqrt{3})J_x\\
-\dfrac{2 |J_x|}{2+\sqrt{3}} < J_z < \dfrac{4|J_x|}{2+\sqrt{3}}
\end{array}\right.
\end{eqnarray}

For any given sublattice, there are only four possible spin orientations, connected between each other by a $\pi/2$ rotation. Any of these four orientations are shared between two different states. In this regard, the $A_{1}$ and $A_{2\perp}$ states are not particular anymore. Every ground-state shares a common spin orientation with three other ground-states. Hence, the stripe order of Fig.~\ref{fig:stripe} remains a possible paving of the lattice. But in addition, diagonal stripes can now co-exist, because crossing triangles between two or three stripes do not cost any energy [Fig.~\ref{fig:cstripe}]; they can be introduced while keeping all triangles within the 8-fold degenerate set of ground-states.

%======================
\begin{figure}[t]
\centering\includegraphics[width=7cm]{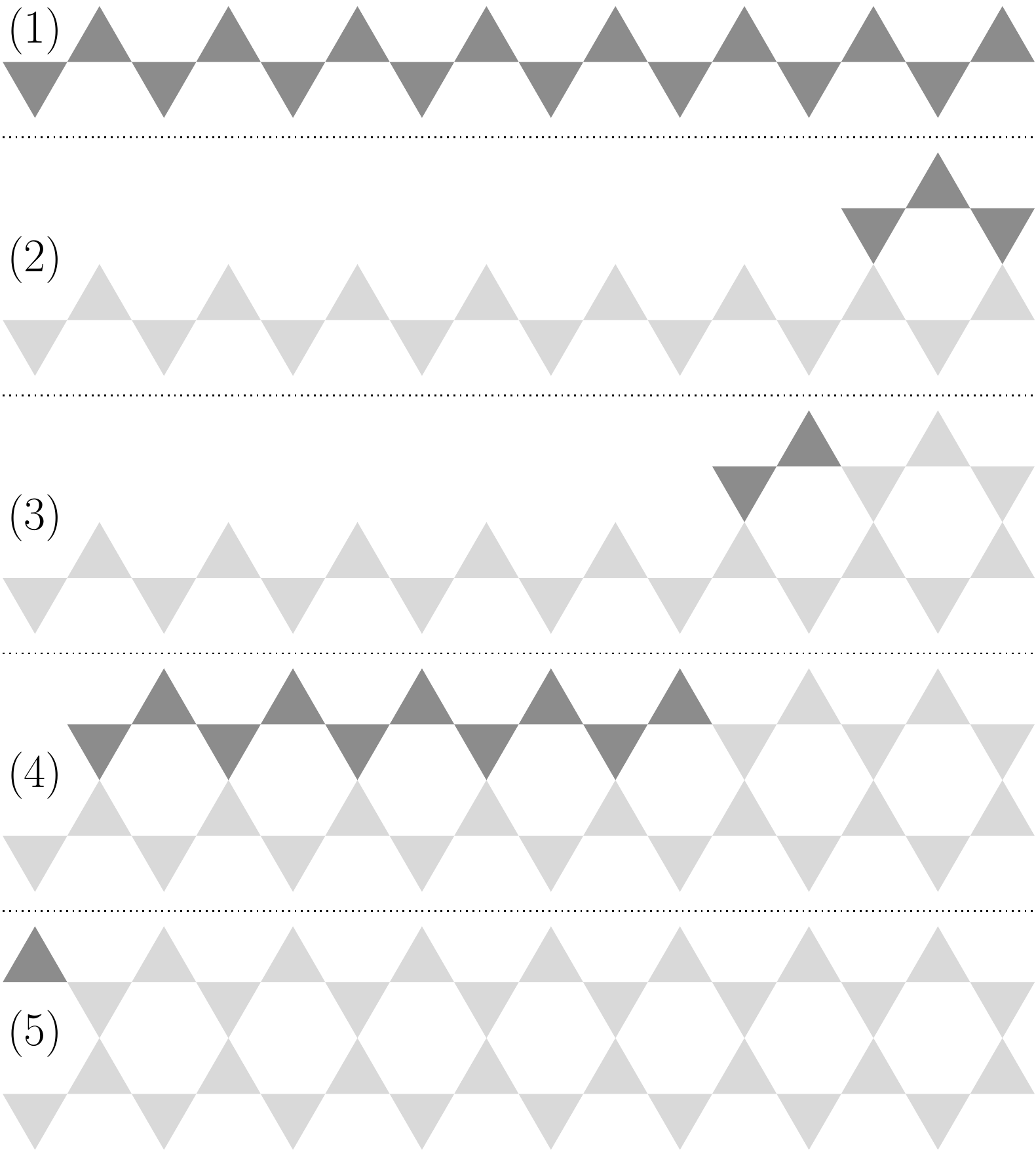}\\
\caption{Five-step illustration on how to calculate the entropy of the ensemble of crossing-stripe configurations. For the sake of clarity, the size of the lattice on step (5) is $L_{x}=16$ and $L_{y}=2$, which gives $N_{\Delta}=L_{x}L_{y}=32$ triangles.
}
\label{fig:cstripeentropy}
\end{figure}
%======================

Even if it is not as straightforward as for the simple stripe order of Fig.~\ref{fig:stripe}, the number of configurations $\Omega_{cs}$ in the crossing-stripe ground-state remains exactly countable. Let us consider a kagome lattice whose borders have a tetragonal shape [as represented in Figs.~\ref{fig:stripe} and Fig.~\ref{fig:cstripe} for example] and open-boundary conditions. There are $L_{x}$ triangles in the horizontal direction and $L_{y}$ lines of triangles in the vertical direction [Fig.~\ref{fig:cstripeentropy}]. Please note that our choice to consider a tetragonal shape of the kagome lattice differentiates $L_{x}$ from $L_{y}$. The counting argument goes as follows [Fig.~\ref{fig:cstripeentropy}]:
\begin{itemize}
\item step (1): Once the first triangle is chosen among the 8 possible ground-states, there are 2 possible choices for each of the remaining triangles on the line: $\omega_{1} = 8\times 2^{L_{x}-1}=4\times 2^{L_{x}}$.
\item step (2): For any spin configuration of the bottom line of triangles, a simple exhaustive counting of possibilities shows that there are always 2 possible choices to add the above three triangles. This can be understood as follows. With the orientation of the bottom spins fixed, there are $2\times 2=4$ possibilities for the two triangles just above the bottom line; to connect these two triangles via the top central triangle eliminates 2 choices, which leaves: $\omega_{2} = 2$.
\item step (3): The addition of the next two triangles is uniquely determined: $\omega_{3} = 1$.
\item step (4): By repeating step (3) until (almost) the end of the line, one gets: $\omega_{4} = 1$.
\item step (5): The last remaining triangle is only constrained by one spin, which always leads to 2 possible choices: $\omega_{5} = 2$.
\end{itemize}
Repeating steps (2-5) for each additional line of triangles in the vertical direction gives the overall number of configurations
\begin{eqnarray}
\Omega_{cs}=\left(4\times 2^{L_{x}}\right) \; 4^{L_{y}-1}=2^{L_{x}+2L_{y}},
\label{eq:omegacs}
\end{eqnarray}
with a sub-extensive ground-state entropy. The same result can be obtained by counting how many stripes can be made on the lattice [see the grey lines in Fig.~\ref{fig:cstripe}]. This is because for a given triangle, there are 8 possible ways to place stripes (or not) around its three edges; it corresponds to the $\mathbb{Z}_{8}$ degeneracy of the ground-states. This proves that any configuration can be obtained from any other configuration by adding a finite number of stripes on the lattice.

%======================================================
\subsubsection{At the frontier with out-of-plane ferromagnetism:\\addition of the $A_{2z}$ irrep}
\label{ssec:A1A2bEminA2a}

As can be deduced from Eqs.(\ref{eq:Z8J1}) and (\ref{eq:Z8J2}), when $J_{z}=-\dfrac{2 |J_y|}{2+\sqrt{3}}$ or $J_{z}=-\dfrac{2 |J_x|}{2+\sqrt{3}}$, we reach models whose ground-states are described by the $E_{min} \oplus A_{2z} \oplus (A_{1}\textrm{ or }A_{2\perp})$ irreps. Since the $A_{2z}$ irrep carries out-of-plane ferromagnetism, the consequences are relatively straightforward.

In the ground-state, the in-plane spin components are described by the $\mathbb{Z}_{8}$ degeneracy of Eqs.~(\ref{eq:ups1}-\ref{eq:ups4}) and (\ref{eq:ups1p}-\ref{eq:ups4p}), while the out-of-plane components take the same value $S_{i=1,3}^{z}=S^{z}\in [-1:+1]$, the ratio between the two being given by normalization $|\mathbf{S}_{i}^{\perp}|^2+(S_{i}^{z})^{2}=1,\quad \forall i=0,1,2$. For each triangle, the degeneracy is now $\mathbb{Z}_{8} \otimes$ O(2).

%======================================================
%======================================================
\subsection{Tricolour spin liquid\\ with local O(2)$\times\mathbb{Z}_{2}$ invariance: $E_{min} \oplus E_{z}$}
\label{sec:tricolor}

%======================================================
\subsubsection{Spin configurations}
\label{ssec:EminEcconfig}

Let us turn our attention to what happens when the $E_{min}$ and $E_{z}$ regions meet. Both irreps correspond to states whose spins are not normalized in length. But luckily, linear combinations of the two provide a manifold of physical states with normalized spins. The procedure is the same as what has been done so far. Using the general expressions of the spins given in Eqs.~(\ref{eq:S0}-\ref{eq:S2}), one imposes that $\mathbf{m}_{I\in\{A_{1},A_{2z},A_{2\perp},E_{max}\}}=0$ and $|S_{i=0,1,2}|^{2}=1$. The spin configurations respecting these conditions are as follows (where $\eta$ has been defined in Eq.~(\ref{eq:eta}))
\begin{widetext}
\begin{align}
\label{eq:O2Z2p}
\eta>0: 
\mathbf{S}_{0}=\left(\begin{array}{ll}
\dfrac{(\eta +2) \cos (s)-\sqrt{3} \eta  \sin (s)}{2 (1+\eta)}\\
-\dfrac{\sqrt{3} \eta  \cos (s)+(\eta -2) \sin (s)}{2 (1+\eta)}\\
\pm\dfrac{2 \sqrt{\eta }}{1+\eta} \sin \left(s+\dfrac{\pi }{6}\right)
\end{array}\right),\;
\mathbf{S}_{1}=\left(\begin{array}{ll}
\dfrac{(\eta +2) \cos (s)+\sqrt{3} \eta  \sin (s)}{2 (1+\eta)}\\
\dfrac{\sqrt{3} \eta  \cos (s)-(\eta-2)  \sin (s)}{2 (1+\eta)}\\
\mp\dfrac{2 \sqrt{\eta }}{1+\eta} \sin \left(s-\dfrac{\pi}{6}\right)
\end{array}\right),\;
\mathbf{S}_{2}=\left(\begin{array}{ll}
\dfrac{1-\eta}{1+\eta}\cos (s)\\
\sin (s)\\
\pm\dfrac{-2 \sqrt{\eta }}{1+\eta} \cos (s)
\end{array}\right)\\
\label{eq:O2Z2n}
\eta<0: 
\mathbf{S}_{0}=\left(\begin{array}{ll}
\dfrac{(\eta +2) \cos (s)-\sqrt{3} \eta  \sin (s)}{2 (1-\eta)}\\
-\dfrac{\sqrt{3} \eta  \cos (s)+(\eta -2) \sin (s)}{2 (1-\eta)}\\
\pm\dfrac{2 \sqrt{-\eta }}{1-\eta} \cos \left(s+\dfrac{\pi }{6}\right)
\end{array}\right),\;
\mathbf{S}_{1}=\left(\begin{array}{ll}
\dfrac{(\eta +2) \cos (s)+\sqrt{3} \eta  \sin (s)}{2 (1-\eta)}\\
\dfrac{\sqrt{3} \eta  \cos (s)-(\eta-2)  \sin (s)}{2 (1-\eta)}\\
\mp\dfrac{2 \sqrt{-\eta }}{1-\eta} \cos \left(s-\dfrac{\pi}{6}\right)
\end{array}\right),\;
\mathbf{S}_{2}=\left(\begin{array}{ll}
\cos (s)\\
\dfrac{1+\eta}{1-\eta}\sin (s)\\
\pm\dfrac{2 \sqrt{-\eta }}{1-\eta} \sin (s)
\end{array}\right)
\end{align}
\end{widetext}
where $s\in[0:2\pi]$ is a O(2) degree of freedom. For a single triangle, the ground-state symmetry is O(2)$\times\mathbb{Z}_{2}$. While the original O(2) invariance of the XXZDM model remains intrinsically broken, a ``deformed'' in-plane O(2) invariance is recovered in the ground-state by allowing variations of the $S^z$ components. It is deformed because the $S^x$ and $S^y$ components are equivalent only up to a prefactor which is function of $\eta$; $S^{x}$ and $S^{y}$ form an ellipsoid upon varying $s$. Since there is no coupling between in-plane and out-of-plane components, the time-reversal symmetry can be further applied to the $S^{z}$ components alone [see the $\pm$ terms in Eqs.~(\ref{eq:O2Z2p}), (\ref{eq:O2Z2n})], giving rise to the additional $\mathbb{Z}_{2}$ degeneracy.

%======================================================
\subsubsection{Long range order with stripes ($\eta\neq \pm1$)}
\label{ssec:LROstripe}

For any set of coupling parameters \{$J_{x},J_{y},J_{z},D$\} there
corresponds a given value of $\eta$ [Eqs.~(\ref{eq:phi}), (\ref{eq:eta})], which gives the spin configurations of Eqs.~(\ref{eq:O2Z2p}) and (\ref{eq:O2Z2n}) for positive and negative $\eta$ respectively. Let us consider the spin $\mathbf{S}_{2}$ without loss of generality. Following the same argument as in the previous sections, in order to have two neighbouring triangles in a different ground-state -- \textit{i.e.} something different from trivial $\mathbf{q}=0$ order -- one needs to find different ground-states sharing at least one spin in common. As mentioned above, the in-plane spin components form an ellipsoid $\mathfrak{E}_{\eta}$ when varying $s$. Since the function
\begin{eqnarray}
\label{eq:aleph}
\aleph_{\eta}: [0:2\pi]\longrightarrow &\hspace{-1cm} \mathfrak{E}_{\eta}\\
\nonumber
s\longmapsto & \left\{
\begin{array}{ll}
\left(\frac{1-\eta}{1+\eta}\cos (s),\sin (s)\right) \textrm{ if } \eta>0\\
\left(\cos (s),\frac{1+\eta}{1-\eta}\sin (s)\right) \textrm{ if } \eta<0
\end{array}\right.
\end{eqnarray}
is bijective for $\eta\neq \pm1$, it means it is not possible to use the O(2) degeneracy to find two different states with at least one spin in common. Before turning our attention to the special cases $\eta=\pm1$, we shall first consider the other, $\mathbb{Z}_{2}$, degeneracy.\\

If all $S_{i=0,1,2}^{z}$ components are finite, then the ground-state is uniquely defined. However, if one of them is zero -- \textit{e.g.} $S_{2}^{z}=0$ for $s=\pm\pi/2$ and $\eta>0$ -- then the $\mathbb{Z}_{2}$ degeneracy ensures two different states with one spin in common, namely $\mathbf{S}_{2}=(0,1,0)$. It is not possible to have two spins in common between different states in this context.

What kind of degeneracy do we obtain ? Since $s$ is fixed, it means the in-plane spin components are long-range ordered, described by a $\mathbf{q}=0$ wavevector. The $\mathbf{S}_{2}$ spin is actually fully ordered since its $S^z$ component is nil. As for $S_{0}^{z}$ and $S_{1}^{z}$, they have $\mathbb{Z}_{2}$ degeneracy but as soon as, say, a $S_{0}^{z}$ component is chosen, its $S_{1}^{z}$ neighbour is fixed, and so on along a line of 0,1,0,1... nearest neighbours. Since the neighbouring line 0,1,0,1... is separated by a row of $\mathbf{S}_{2}$ spins, one gets the same kind of stripe order as depicted in Fig.~\ref{fig:stripe} but where the one-dimensional disordered degree of freedom is the $S^{z}$ component of the 0 and 1 sublattices. By choosing different values of $s$, \textit{e.g.} $\{-\frac{\pi}{6},\frac{5\pi}{6}\}$ or $\{\frac{\pi}{6},\frac{7\pi}{6}\}$ for $\eta>0$, the stripes can be made diagonal.

%======================================================
\subsubsection{Tricolour spin liquids ($\eta = \pm1$)}
\label{ssec:tricolour}

For $\eta=\pm 1$, the function $\aleph_{\eta}$ of Eq.~(\ref{eq:aleph}) is not bijective anymore, but remains surjective. The O(2) degeneracy can now be exploited to 
allow more non-trivial tilings of the lattice. The spin configurations become
\begin{widetext}
\begin{align}
\label{eq:O2Z2p1}
\Psi_{\eta=+1}^{\pm}(s)=\left\{
\mathbf{S}_{0}=\left(\begin{array}{ll}
\cos \left(-\dfrac{\pi }{6}\right) \cos\left(s+\dfrac{\pi }{6}\right)\\
\sin \left(-\dfrac{\pi }{6}\right) \cos\left(s+\dfrac{\pi }{6}\right)\\
\pm \sin \left(s+\dfrac{\pi }{6}\right)
\end{array}\right),\;
\mathbf{S}_{1}=\left(\begin{array}{ll}
\cos \left(\dfrac{\pi }{6}\right) \cos\left(s-\dfrac{\pi }{6}\right)\\
\sin \left(\dfrac{\pi }{6}\right) \cos\left(s-\dfrac{\pi }{6}\right)\\
\mp \sin \left(s-\dfrac{\pi}{6}\right)
\end{array}\right),\;
\mathbf{S}_{2}=\left(\begin{array}{ll}
0\\\sin (s)\\
\mp \cos (s)
\end{array}\right)
\right\},\\
\label{eq:O2Z2n1}
\Psi_{\eta=-1}^{\pm}(s)=\left\{
\mathbf{S}_{0}=\left(\begin{array}{ll}
\sin \left(\dfrac{\pi }{6}\right)  \sin\left(s+\dfrac{\pi }{6}\right)\\
\cos \left(\dfrac{\pi }{6}\right)  \sin\left(s+\dfrac{\pi }{6}\right)\\
\pm \cos \left(s+\dfrac{\pi }{6}\right)
\end{array}\right),\;
\mathbf{S}_{1}=\left(\begin{array}{ll}
\sin \left(-\dfrac{\pi }{6}\right)  \sin\left(s-\dfrac{\pi }{6}\right)\\
\cos \left(-\dfrac{\pi }{6}\right)  \sin\left(s-\dfrac{\pi }{6}\right)\\
\mp \cos \left(s-\dfrac{\pi}{6}\right)
\end{array}\right),\;
\mathbf{S}_{2}=\left(\begin{array}{ll}
\cos (s)\\
0\\
\pm \sin (s)
\end{array}\right)
\right\},
\end{align}
\end{widetext}
%

%======================
\begin{figure}[b]
\centering\includegraphics[width=7cm]{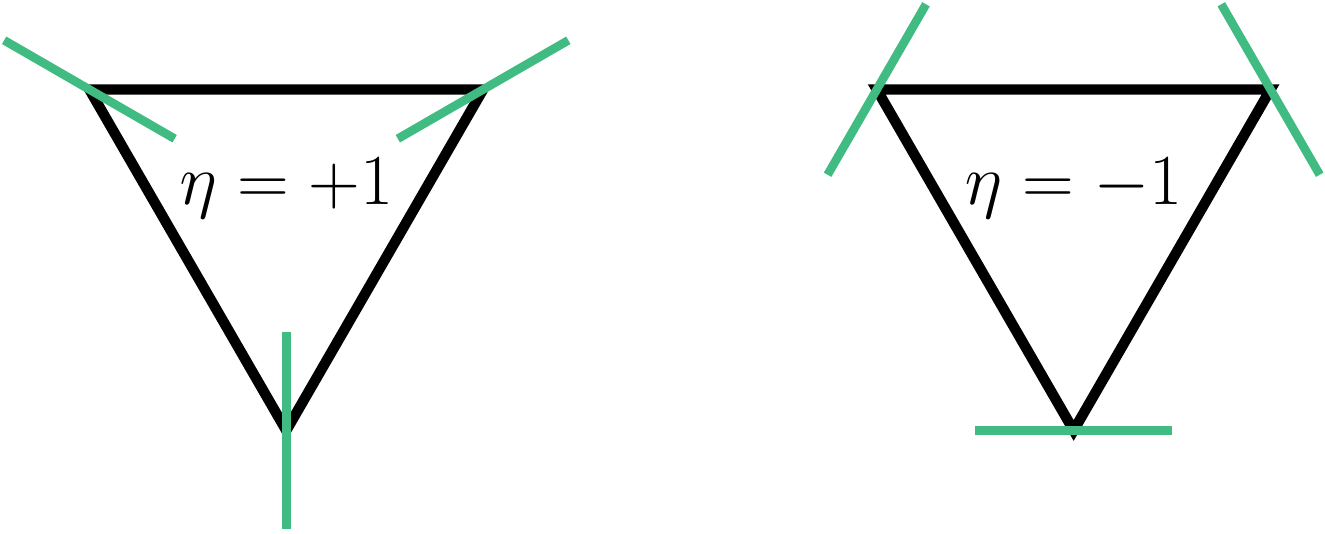}
\caption{Projection (in green) of the ground-state configurations in the kagome plane, at the frontier between $E_{min}$ and $E_{z}$, for $\eta=\pm1$ [Eqs.~(\ref{eq:O2Z2p1}) and (\ref{eq:O2Z2n1})]. When expressed in the local bases $\mathcal{B}^{k=0,1,2}$ of Fig.~\ref{fig:basisB}, the O(2) invariance takes the form of a circle in spin space which lies entirely in the local $(y,z)_{k=0,1,2}$ ($\eta=+1$, left) or $(x,z)_{k=0,1,2}$ ($\eta=-1$, right) planes}
\label{fig:O2eta1}
\end{figure}
%======================
%
as illustrated in Fig.~\ref{fig:O2eta1}. In order to determine how to connect ground-state configurations next to each other (via at least one spin in common), the idea is to
\begin{enumerate}[i.]
\item randomly choose a ground-state labeled $\ell$ and a sublattice $k=\{0,1,2\}$,
\item flip the sign of the $S_{z}$ components thanks to the $\mathbb{Z}_{2}$ degeneracy,
\item use the O(2) degeneracy parametrised by $s$ to recover the same spin on sublattice $k$ as in the ground-state $\ell$; this transformation is unique and gives a new ground-state $\ell+1$. Let us randomly choose a new sublattice $k'\neq k$,
\item repeat steps (ii) and (iii) until a closed set is obtained, \textit{i.e.} that further iterations reproduce only ground-states of the set.
\end{enumerate}
%
%======================
\begin{figure*}
\centering{\Large 
$\eta=+1$ \includegraphics[width=18cm]{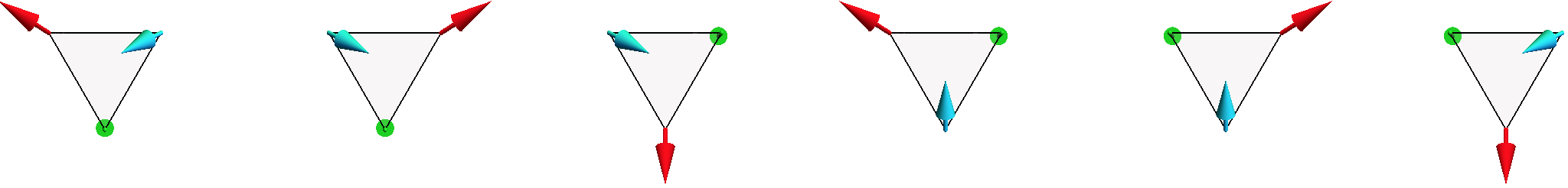}
$\eta=- 1$ \includegraphics[width=18cm]{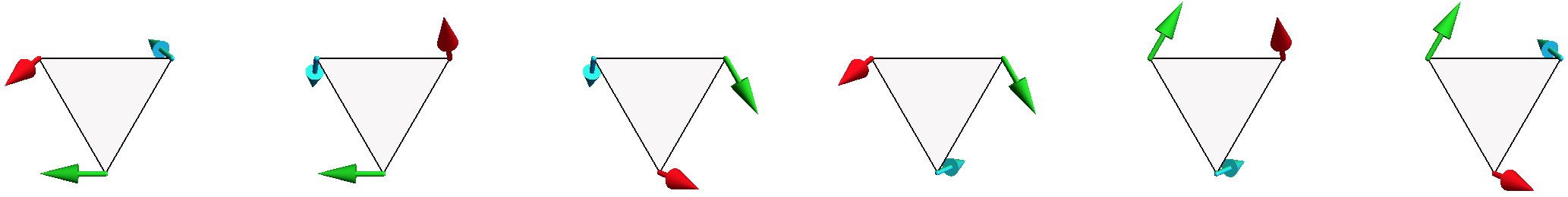}}
\caption{Sextets of ground-states at the frontier between $E_{min}$ and $E_{z}$ for $\eta=+1$ (top) and $\eta=-1$ (bottom) [Eqs.~(\ref{eq:O2Z2p1}) -- (\ref{eq:sextet-1})]. Each spin orientation appears in two different ground-states. The paving of the kagome lattice by these six states is equivalent to a tricolouring problem, and the spins have been coloured accordingly. Here $s$ has been arbitrarily fixed to a value of $\pi$.}
\label{fig:tricolour}
\end{figure*}
%======================
%
The above procedure gives the following sextets of spin configurations [$\Psi_{\pm}^{\pm}(s)$ has been defined in Eqs.~(\ref{eq:O2Z2p1},\ref{eq:O2Z2n1})]
\begin{eqnarray}
\eta=+1:
\left\{\begin{array}{ll}
\Psi_{+1}^{\pm}\left(s\right),&\Psi_{+1}^{\mp}\left(\pi-s\right)\\
\Psi_{+1}^{\pm}\left(\dfrac{2\pi}{3}+s\right),&\Psi_{+1}^{\mp}\left(\dfrac{5\pi}{3}-s\right)\\
\Psi_{+1}^{\pm}\left(\dfrac{4\pi}{3}+s\right),&\Psi_{+1}^{\mp}\left(\dfrac{7\pi}{3}-s\right)
\end{array}\right\}
\label{eq:sextet+1}
\end{eqnarray}
\begin{eqnarray}
\eta=-1:
\left\{\begin{array}{ll}
\Psi_{-1}^{\pm}\left(s\right),&\Psi_{-1}^{\mp}\left(-s\right)\\
\Psi_{-1}^{\pm}\left(\dfrac{2\pi}{3}+s\right),&\Psi_{-1}^{\mp}\left(\dfrac{2\pi}{3}-s\right)\\
\Psi_{-1}^{\pm}\left(\dfrac{4\pi}{3}+s\right),&\Psi_{-1}^{\mp}\left(\dfrac{4\pi}{3}-s\right)
\end{array}\right\}
\label{eq:sextet-1}
\end{eqnarray}
for any value of $s\in[0:2\pi/3[$. There is no need to consider further values of $s$ since there is a $2\pi/3$ periodicity in the sextets. Remarkably, for each sextet, when the spins on each sublattice $k$ are expressed in the local bases $\mathcal{B}^{k=0,1,2}$ [Fig.~\ref{fig:basisB}], they correspond to only three different orientations. 
This means that each sextet can be mapped onto a tricolouring of the triangle, as illustrated in Fig.~\ref{fig:tricolour} for $s=\pi$. Such tricolouring paving is possible on the kagome lattice and bears a countable and extensive entropy \cite{Baxter1970}. It is known to describe the ground-state of the XXZ [\onlinecite{Huse1992}], and equivalent XXZ$^{\pm}$ [\onlinecite{Essafi16a}] Hamiltonians [section \ref{sec:XXZDMSL}].

However, the tricolouring paving of the present ground-state carries an additional property. Since the spins are not coplanar, it means that the ground-states may carry a finite scalar chirality. Following the definition of Eq.~(\ref{eq:scalarchiral}), one obtains
\begin{eqnarray}
\kappa\left[\Psi_{+1}^{\pm}(s)\right]=\mp \dfrac{3\sqrt{3}}{8}\cos(3s)\\
\kappa\left[\Psi_{-1}^{\pm}(s)\right]=\mp \dfrac{3\sqrt{3}}{8}\sin(3s)
\label{eq:scalarchiraltricolour}
\end{eqnarray}
One can easily check that all ground-states belonging to a given sextet carry the same scalar chirality. This means that the models at the frontier between the $E_{z}$ and $E_{min}$ irreps for $\eta=\pm 1$ possess an extensively degenerate ground-state with a uniform scalar chirality; this is the classical analogue of a chiral spin liquid.

Please note that in general, for a given tricolour problem on kagome, if each colour were to correspond to the same spin orientation expressed in a \textit{global} frame, then it is not possible to get a finite scalar chirality after statistical average. The intrinsic property of a tricolouring problem is that, for each triangle, permuting any pair of spins remains a valid configuration. Since this permutation reverses the sign of the scalar chirality, averaging over the ensemble of tricolour states necessarily gives zero scalar chirality. Hence a necessary, but not sufficient, condition for chiral tricolour spin liquids is for the colouring to correspond to the same spin orientations in different \textit{local} frames.\\

To conclude this section, one needs to provide the Hamiltonians -- \textit{i.e.} the values of $\{J_{x},J_{y},J_{z},D\}$ -- supporting such classical chiral spin liquids as their ground-states. Using Eqs.(\ref{eq:phi}) and (\ref{eq:eta}) for $\eta=\pm 1$, one obtains
\begin{eqnarray}
\label{eq:tricolour1}\left\{
\begin{array}{ll}
D=\frac{\sqrt{3}}{2} \left(J_x+J_y\right)\\
J_z =\frac{1}{2} \left(-3 J_x-J_y\right)\\
J_{x}<J_{y}<-\frac{5}{7}J_{x}
\end{array}\right.\\
\label{eq:tricolour2}\left\{
\begin{array}{ll}
D=\frac{\sqrt{3}}{2} \left(J_x+J_y\right)\\
J_z =\frac{1}{2} \left(-3 J_y-J_x\right)\\
J_{y}<J_{x}<-\frac{5}{7}J_{y}
\end{array}\right.
\end{eqnarray}
When $J_{x}=J_{y}<0$, one recovers a special point of the XXZDM model, equivalent of the Heisenberg antiferromagnet, where the $E_{FM}$, $E_{AF}$ and $E_{z}$ states are all degenerate in the ground-state (see Table \ref{tab:SL}). On the other hand, when $J_{y}=-\frac{5}{7}J_{x}>0$ or $J_{x}=-\frac{5}{7}J_{y}>0$, the $E_{min}$ and $E_{z}$ irreps meet respectively the $A_{1}$ and $A_{2\perp}$ irreps in the ground-state. Such models are expected to possess a very high degeneracy at zero temperature, but the exact, and complete, determination of their ground-state manifolds becomes challenging.

%======================================================
%======================================================
%======================================================
\section{Comparison between\\kagome \& pyrochlore}
\label{sec:Mach}

%======================
\begin{figure*}[t]
\centering\includegraphics[width=17cm]{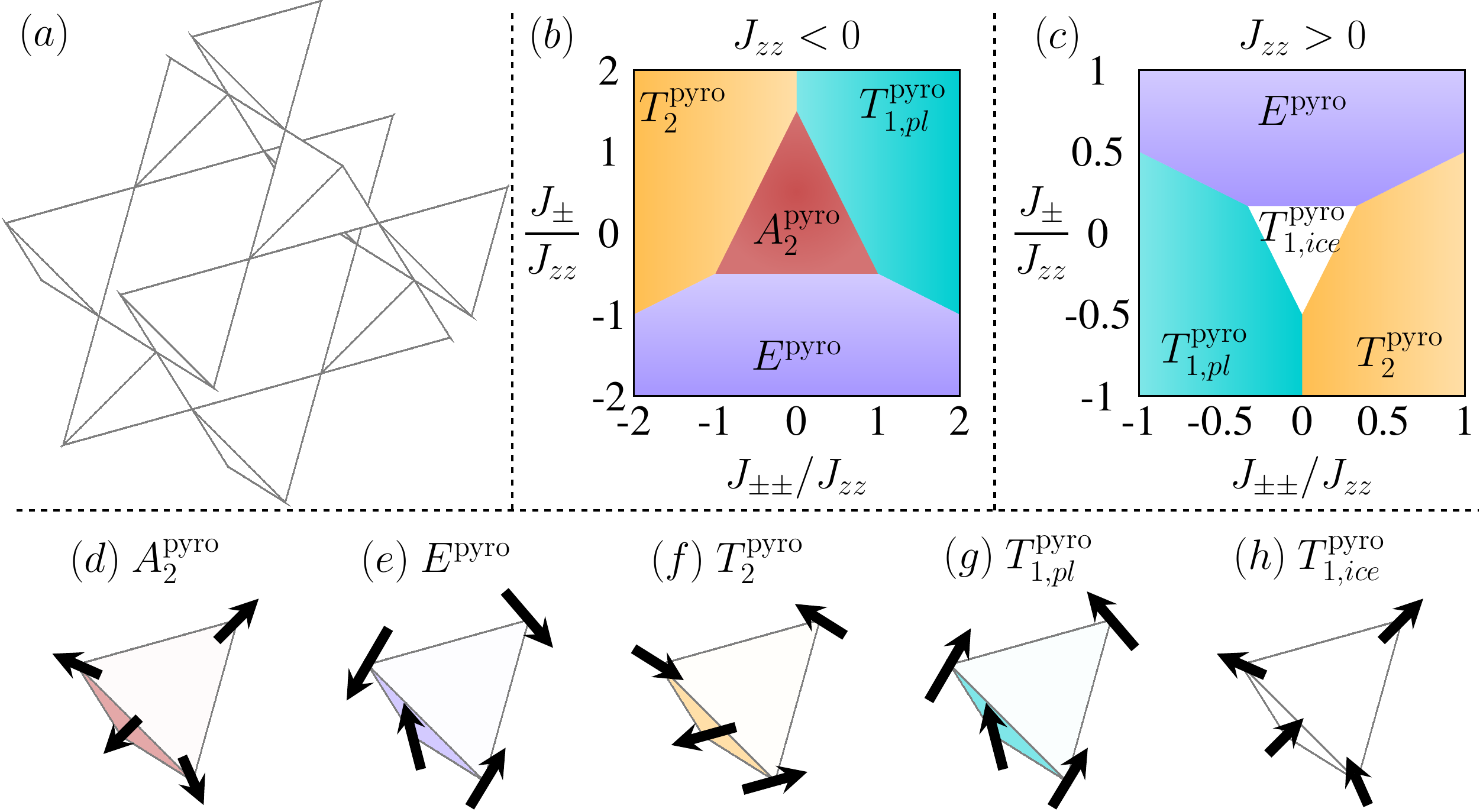}
\caption{$(a)$ Pyrochlore lattice whose minimal unit cell is a tetrahedron, made of four sublattices. Each sublattice possesses an easy axis connecting the centre of the two neighbouring tetrahedra. The spin component along this easy axis is noted $S^{z}$. 
$(b,c)$ Ground states of Hamiltonian (\ref{eq:Hqsi}) for classical Heisenberg spins assuming a negative $(b)$ and positive $(c)$ value of $J_{zz}$. Panel $(b)$ is derived from the results of Ref.~[\onlinecite{Yan17a}], together with the irrep decomposition illustrated in panels $(d-h)$. Panel $(c)$ has been obtained in Ref.~[\onlinecite{Taillefumier17a}], while semi-classical and quantum mean field versions thereof were derived in Refs.~[\onlinecite{Onoda11a},\onlinecite{Petit16b}] and Ref.~[\onlinecite{Lee12a}] respectively. 
$(d)$ ``all in all out'' states transforming like the $A_{2}^{\rm pyro}$ irrep.
$(e)$ $\Gamma_{5}$ states \cite{Poole07a} transforming like the $E^{\rm pyro}$ irrep.
$(f)$ Palmer-Chalker states \cite{Palmer00a} transforming like the $T_{2}^{\rm pyro}$ irrep.
$(g)$ Canted, easy plane, ferromagnetic states transforming like the $T_{1}^{\rm pyro}$ irrep. 
We label the basis vector corresponding to this state as  $T_{1 pl}^{\rm pyro}$.
$(h)$ Spin-ice state, with two spins pointing inside and two pointing outside the tetrahedron, which transforms like the $T_{1}^{\rm pyro}$ irrep. 
We label the basis vector corresponding to this state as  $T_{1 ice}^{\rm pyro}$.
Spins configurations in panels (d,h) and (e,g,h) lie respectively along their local easy axes and within their local easy planes. All results in this figure can be found in Refs.~[\onlinecite{Onoda11a,Benton2014,Petit16b,Yan17a,Taillefumier17a}].
}
\label{fig:pyro}
\end{figure*}
%======================

When compared to pyrochlores, the emergence of states with non-normalised spins is rather remarkable. When the lowest eigenvalue
\begin{eqnarray}
\lambda_{min}=\min\{\lambda_{A_1},\lambda_{A_2, z},\lambda_{A_2, \perp},\lambda_{E,\alpha},\lambda_{E,\beta},\lambda_{E, z}\} \quad
\label{eq:loweigen}
\end{eqnarray}
is not degenerate and corresponds to $E_{\alpha}, E_{\beta}$ or $E_{z}$, the ground-state necessarily includes different kinds of magnetic order. In other words more than one order parameter, as defined in Eqs.~(\ref{eq:Order-Parameter-A1}-\ref{eq:Order-Parameter-Ec}), is non zero.

A similar irrep decomposition to the one done in this paper has been made on the pyrochlore lattice \cite{Yan17a} for the generic nearest-neighbour Hamiltonian \cite{Curnoe2007} [Fig.~\ref{fig:pyro}]. In pyrochlores, all irreps correspond to configurations with normalised spins. Unless the lowest eigenvalue $\lambda_{min}$ is degenerate, there is no co-existence of magnetic order in the classical ground-state. This is a strong qualitative difference between two of the most studied lattices in frustrated magnetism.

One should understand that the presence of the $E_{\alpha,\beta,z}$ irreps on kagome is not an artefact of the method. For both kagome and pyrochlore, the Hamiltonians have been derived solely based on the symmetries of the lattice, and then diagonalised for the minimal unit cell. The resulting eigenbasis corresponds to physical spin configurations on pyrochlore, while on kagome it does not. This can be intuitively rationalised as follows, where the irreps and basis vectors will be labeled by ``pyro'' or ``kag'' for clarity.\\

The pyrochlore lattice possesses a cubic symmetry which means that the $x,y$ and $z$ axes are equivalent. However, for each sublattice, there is a given easy axis which defines a local $S^{z}$ components [Fig.~\ref{fig:pyro} and Appendix \ref{app:coordinates}]. This easy axis on pyrochlore plays a role similar to the global $z$-axis on kagome. For example the ``ferromagnetic'' state where $S^{z}=1$ for all spins transforms like the $A_{2}$ irrep on both lattices \cite{Yan17a}. It corresponds to the ``all in all out'' state on pyrochlore [Fig.~\ref{fig:pyro}.$(d)$]. One needs to keep in mind that for pyrochlore, $S^{z}$ is defined in a local frame; a ferromagnetic state expressed in the local frame is actually antiferromagnetic in the global one. On kagome (resp. pyrochlore), the only other basis vector with $S^{z}$ components is $E_{z}^{kag}$ (resp. $T_{1 ice}^{\rm pyro}$, see Fig.~\ref{fig:pyro}.$(h)$). These basis vectors are ``antiferromagnetic'' in the sense that $\sum_{i}S_{i}^{z}=0$. This is a necessity because the ``ferromagnetic'' contribution has already been accounted for in the $A_{2}$ basis vectors. Hence, the non-normalised spins of $E_{z}^{kag}$ comes from the trivial fact that the sum of three Ising degrees of freedom cannot be zero. This is of course possible for four spins on a tetrahedron, which is why all spins in the $T_{1, ice}^{\rm pyro}$ basis vector are normalised.\\

Here we have an interesting analogy between the $E_{z}^{kag}$ phase on kagome and the spin-ice physics supported by the $T_{1, ice}^{\rm pyro}$ irrep. The analogy can be made quantitative when applied to a special case of the generic nearest-neighbour Hamiltonian on pyrochlore \cite{Curnoe2007,Onoda10a,Onoda11a,Ross2011,Lee12a,Petit16b}
\begin{eqnarray}
\mathcal{H}_{\sf QSI} = \sum_{\langle ij\rangle} &
J_{zz}\, S^{z}_i S^{z}_j 
- J_\pm \left( S^+_i S^-_j + S^-_iS^+_j \right) \nonumber\\
+&J_{\pm\pm} \left[\gamma_{ij} S_i^+ S_j^+ + \gamma_{ij}^*
                 S_i^-S_j^-\right] \; ,
    \label{eq:Hqsi}
\end{eqnarray}
where the classical Heisenberg spins are expressed in their local frames, with $S^\pm=S^x\pm \imath S^y$, and $\gamma_{ij}$ are complex phase factors; see Appendix \ref{app:coordinates} for the definitions of the local frames and $\gamma_{ij}$. The classical ground-states of Hamiltonian (\ref{eq:Hqsi}) are known \cite{Onoda11a,Benton2014,Petit16b,Yan17a,Taillefumier17a} and reproduced in Fig.~\ref{fig:pyro}.$(b,c)$. This phase diagram on pyrochlore displays the same form as the XXZDM model on kagome [Fig.~\ref{fig:PhD}.(c)], by replacing $J_{\pm\pm}$ with the Dzyaloshinskii-Moriya coupling $D$. On their respective models, the $T_{1, ice}^{\rm pyro}$ and $E_{z}^{kag}$ irreps have minimal eigenvalue over a triangular region, surrounded by three long-range ordered phases whose spins lie in their easy- or kagome planes. For $J_{zz}$ negative, the same description holds with the $A_{2}$ irreps sitting in the middle. One difference though is that on kagome, the three surrounding ``in-plane'' irreps are two dimensional, while on pyrochlore they are either two-dimensional ($E^{\rm pyro}$) or three-dimensional ($T_{2}^{\rm pyro}$ and 
$T_{1,pl}^{\rm pyro}$). It means that the three-fold symmetry of the phase diagram on kagome [see Eq.~(\ref{eq:EaEbA}) and Ref.~[\onlinecite{Essafi16a}]] is only two-fold on pyrochlore
\begin{eqnarray}
J_{\pm\pm} \longleftrightarrow -J_{\pm\pm} \quad\&\quad T_{2}^{\rm pyro} \longleftrightarrow T_{1,pl}^{\rm pyro}.
\label{eq:T2T1pl}
\end{eqnarray}
with an isosceles, rather than equilateral, triangular region in the middle of the phase diagrams of Fig.~\ref{fig:pyro}.$(b,c)$.\\ 

From this point of view, the Ising antiferromagnet, residing at the centre of the $E_{z}^{kag}$ white triangle of Fig.~\ref{fig:PhD}.(c), is the kagome analogue of spin ice. The reason why the physics of these two models is qualitatively different largely stems from the non-normalised spins in the $E_{z}^{kag}$ irrep. The analogies, and differences, between the $E_{z}^{kag}$ and $T_{1, ice}^{\rm pyro}$ irreps are a vivid illustration of what happens between the kagome and pyrochlore lattices on a broader scale.\\

Indeed, our kagome/pyrochlore comparison has so far been restricted to the XXZDM model and the Hamiltonian of Eq.~(\ref{eq:Hqsi}). As discussed in this paper, the kagome symmetry allows for the XXZDM model to lose its in-plane O(2) invariance and to become the XYZDM model. An important consequence is that the Hamiltonian diagonalisation then requires the mixing of the $E_{FM}^{\rm kag}$ and $E_{AF}^{\rm kag}$ irreps into the non-normalised $E_{\alpha}^{\rm kag}$ and $E_{\beta}^{\rm kag}$ eigenstates [Eq.~(\ref{eq:malpha-mbeta})]. Similarly, the pyrochlore symmetry allows for a more generic Hamiltonian than the one of Eq.~(\ref{eq:Hqsi}) [\onlinecite{Curnoe2007}]. In pyrochlore, the additional interaction takes the form of a coupling between the easy-plane and easy-axis spin components, $J_{z\pm}$ [\onlinecite{Ross2011}]. This coupling also induces a mixing between the states transforming according to the $T^{\rm pyro}_1$ irrep, namely 
$T_{1,ice}^{\rm pyro}$ and $T_{1,pl}^{\rm pyro}$. However, the resulting eigenvectors remain physical in the sense that all spins are normalised [\onlinecite{Yan17a}]. 
\\

On pyrochlore, the even and larger number of spins in the minimal unit cell makes it easier $(i)$ to accommodate frustration and $(ii)$ to support a variety of classical spin liquids \cite{Harris1997,Moessner1998,Canals01a,Benton16a,Yan17a,Taillefumier17a} thanks to linear combination of ground-state irreps. This propensity of the pyrochlore lattice for spin liquids is consistent with the Moessner-Chalker criterion for O($n$) antiferromagnets \cite{Moessner1998} [see Appendix \ref{app:MCcriterion}].

On kagome, the odd and smaller number of spins in the minimal unit cell is responsible for extended regions of parameter space where multiple types of order have to co-exist in the ground-state. Disorder is not necessarily less favoured on kagome, since the Mermin-Wagner-Hohenberg theorem prevents finite-temperature symmetry breaking for a variety of high-symmetry models with Goldstone modes. 
However, the presence of the $E_{\alpha}, E_{\beta}$ and $E_{z}$ irreps induces exotic ordered and disordered phases, as exemplified at the frontier of these regions in section \ref{sec:XYZDM}.

Colloquially speaking, frustration on kagome is more ``pathological'' than on pyrochlore.

%======================================================
%======================================================
%======================================================
\section{Conclusion}
\label{sec:discuss}

The generic nearest-neighbour Hamiltonian allowed by the symmetry of the kagome lattice, where the kagome plane is a mirror plane, is the XYZ model with Dzyaloshinskii-Moriya interactions [Eqs.~(\ref{eq:Hamiltonian},\ref{eq:JbasisB})], described by four coupling parameters $(J_{x},J_{y},J_{z},D)$. The XYZ interactions are properly defined in a set of \textit{local} bases [Fig.~\ref{fig:basisB}]; a simple XYZ Hamiltonian, where the spins would be expressed in the same global basis, is forbidden by kagome symmetry [Eqs.~(\ref{eq:J01}-\ref{eq:J20})]. As for the Dzyaloshinskii-Moriya vector $\mathbf{D}$, it points out of plane \cite{Elhajal2002}.

Using a decomposition in irreducible representations, the XYZDM Hamiltonian for classical Heisenberg spins can be diagonalised for each triangle. It is quadratic in terms of the order parameters [Eq.~(\ref{eq:hamQ})], which allows for a systematic determination of the ground-state for a broad region of parameter space $(J_{x},J_{y},J_{z},D)$. The picture that emerges is a connected map of ordered phases and classical spin liquids [Fig.~\ref{fig:PhD} and Table \ref{tab:SL}]. In particular, in the XXZ model with Dzyaloshinskii-Moriya, the irrep decomposition sheds a new light on the mapping between three different spin liquids observed in Ref.~[\onlinecite{Essafi16a}]; the spin-liquids ground-states are equivalent up to a permutation of their irreps [Table \ref{tab:SL}]. The $E_{AF}$ and $A$ irreps responsible for the tricolouring spin liquids persist in the ground-state of the $E_{z}$ region [section \ref{ssec:XXZEc}]. This co-existence of phases at the classical level might play a role in the stability of the quantum spin liquid along the XXZ line for quantum sins $S=1/2$ \cite{Lauchli15a,He15a,He2015}.

The XYZDM model is ``asymmetric'' with respect to spin chirality [section \ref{sec:chiasym}]. This enables to energetically differentiate the two states with negative chirality ($\kappa_{z}=-1$), namely $A_{1}$ and $A_{2\perp}$. At finite temperature, this is expected to change the universality class of the phase transition into these ordered states to Ising. This chiral asymmetry also mixes the $E_{AF}$ states with positive chirality ($\kappa_{z}=+1$) together with the in-plane ferromagnetic states, $E_{FM}$. This mixing produces new forms of order and spin liquids. It is possible to stabilise an 8-fold degenerate ground-state at the level of each triangle [Eqs.~(\ref{eq:ups1}-\ref{eq:ups4p}), Figs.~\ref{fig:Z8} and \ref{fig:Z8star}]. This local $\mathbb{Z}_{8}$ degeneracy leads to a global sub-extensive entropy and paves the lattice to form stripe orders, with or without crossings [Fig.~\ref{fig:stripe} and \ref{fig:cstripe}]. In addition to sub-extensive stripe order, there exists a range of coupling parameters [Eqs.~(\ref{eq:tricolour1},\ref{eq:tricolour2})] whose ground-state corresponds to a tricolouring of the kagome lattice [Fig.~\ref{fig:tricolour}]. The colouring corresponds to a different spin orientation depending on the sublattice, which allows for this extensively degenerate ensemble of ground-states to bear a global finite scalar chirality. In other words, this family of models supports a classical (tricolour) chiral spin liquid.

To conclude, in section \ref{sec:Mach}, we have compared the generic models on two of the most studied frustrated lattices: kagome \& pyrochlore. Despite striking analogies on their phase diagrams [see Figs.~\ref{fig:PhD} and \ref{fig:pyro}], the two models differ on the qualitative nature of their irreps, since eigenstates of the generic model always have normalised spins on pyrochlore, but not on kagome -- $E_{\alpha}, E_{\beta}$ and $E_{z}$ irreps. As a consequence, kagome materials can naturally support low-temperature phases with multiple kinds of orders, even without quantum superposition of states or formations of domains. Disordered magnetic textures can also co-exist with long-range order, and be responsible for persistent dynamics below ordering transitions.\\

In this paper, we have provided a detailed exploration of exact results for the generic kagome model with classical Heisenberg spins. We believe this opens several directions of investigation. For example, we have not looked in detail inside the pathological $E_{\alpha}, E_{\beta}$ and $E_{z}$ irreps, nor have we studied the more generic Hamiltonian where kagome plane symmetry is broken [Eqs.~(\ref{eq:J01gen}) - (\ref{eq:J20gen})]. These regions and Hamiltonians very probably hide a richness of exotic phases and unconventional dynamics, where co-existence between order and disorder might be the norm rather than the exception.

Such co-existence is reminiscent of the partial order observed in Vesignieite \cite{quilliam11b} and purified Edwardsite \cite{Fujihala14a} compounds. While lattice distortion has been suggested to be the source of partial order in the latter material, nearest-neighbour anisotropic coupling -- via in-plane $D_{y}$ Dzyaloshinskii-Moriya interactions -- might be responsible for the observed competition between order and disorder in Vesignieite\cite{quilliam11b,yoshida13a,Zorko13a}. The two-step ordering observed in Vesignieite \cite{yoshida13a} would also be consistent with the co-existence of different kinds of order. More generally, rare-earth-based materials such as tripod kagome~\cite{Dun16a,Scheie16a,Paddison16b,Dun17a} offer the strong spin-orbit coupling necessary for highly anisotropic interactions. For comparison to experiments, a study of the finite-temperature properties of the generic XYZDM model would be helpful. The melting of three-sublattice order is for example a famously complex mechanism \cite{Damle15a}. In light of the diverse regions of (sub-)extensive degeneracy, order-by-disorder, multi-step ordering and Berezinsky-Kosterlitz-Thouless transitions are to be expected.

And of course, a large portion of the parameter space forming the XYZDM model is an unexplored territory with quantum spins. The anisotropy of the XYZDM model is a perfect ingredient for the emergence of chiral phases, and the known results for the XXZ quantum spin liquids penetrating the $E_{z}$ region \cite{Lauchli15a,He15a,He2015,Zhu15a,Hu15a,Lauchli16a} makes it exciting to study how quantum fluctuations will mix states that are already co-existing at the classical level \cite{Changlani17a}.

On a more academic level, the comparison between the kagome and pyrochlore lattices raises the question of what happens for the equivalent lattice in four dimensions, made of corner-sharing pentachorons -- the four-dimensional analogues of tetrahedra in 3D and triangles in 2D. In this case, the odd number of spins in the minimal unit cell ($q=5$) comes together with a high number of degrees of freedom to support the stability of disordered phases in the ground-states.

%======================================================
%======================================================
\begin{acknowledgments}
We thank John Chalker, Zhiling Dun, Andreas Laeuchli, Claire Lhuillier \& Laura Messio for useful discussions, as well as Nic Shannon, Mathieu Taillefumier and Han Yan for collaborations on related topics. This work was supported in part by the Okinawa Institute of Science and Technology Graduate University.
\end{acknowledgments}
%======================================================
%======================================================
%======================================================
\begin{appendix}
%======================================================
%======================================================
\section{Definitions of local coordinate frames for the pyrochlore lattice}
\label{app:coordinates}

With respect to the global cubic coordinate frame, the positions of the four spins in a tetrahedron ${\bf S}_0$, ${\bf S}_1$, ${\bf S}_2$, ${\bf S}_3$ are
\begin{eqnarray}
&{\bf r}_0 =   \left( 1, 1, 1 \right) 
\qquad 
&{\bf r}_1 =   \left( 1,-1,-1 \right) 
\nonumber \\
&{\bf r}_2 =   \left( -1,1,-1 \right) 
\qquad 
&{\bf r}_3 =  \left( -1,-1,1 \right).
\label{eq:r}
\end{eqnarray}
For each sublattice, the local easy axes are
\begin{eqnarray}
&\mathbf{z}^{\sf local}_0 = \dfrac{1}{\sqrt{3}}(1,1,1)
\qquad
&\mathbf{z}^{\sf local}_1 = \dfrac{1}{\sqrt{3}}(1,-1,-1) 
\nonumber \\
&\mathbf{z}^{\sf local}_2 = \dfrac{1}{\sqrt{3}}(-1,1,-1)
\qquad
&\mathbf{z}^{\sf local}_3 = \dfrac{1}{\sqrt{3}}(-1,-1,1) 
\; , \nonumber
\label{eq:local-111-axis}\\
\end{eqnarray}
while the easy planes are defined by the local $x-$ and $y-$axes.
\begin{eqnarray}
&\mathbf{x}^{\sf local}_0 = \dfrac{1}{\sqrt{6}}(-2,1,1)
\qquad
&\mathbf{x}^{\sf local}_1 = \dfrac{1}{\sqrt{6}}(-2,-1,-1)
\nonumber \\
&\mathbf{x}^{\sf local}_2 = \dfrac{1}{\sqrt{6}}(2,1,-1)
\qquad
&\mathbf{x}^{\sf local}_3 = \dfrac{1}{\sqrt{6}}(2,-1,1)
\; , \nonumber
\label{eq:local-easy-plane-x}\\
\end{eqnarray}
\begin{eqnarray}
&\mathbf{y}^{\sf local}_0 = \dfrac{1}{\sqrt{2}}(0,-1,1)
\qquad
&\mathbf{y}^{\sf local}_1 = \dfrac{1}{\sqrt{2}}(0,1,-1)
\nonumber \\
&\mathbf{y}^{\sf local}_2 = \dfrac{1}{\sqrt{2}}(0,-1,-1)
\qquad
&\mathbf{y}^{\sf local}_3 = \dfrac{1}{\sqrt{2}}(0,1,1)
\; . \nonumber
\label{eq:local-easy-plane-y}\\
\end{eqnarray}

These local coordinate frames are responsible for complex phase factors in the Hamiltonian of Eq.~(\ref{eq:Hqsi}) [\onlinecite{Ross2011,Savary12a}], defined by a $4\times 4$ matrix
\begin{eqnarray}
\gamma=
\begin{pmatrix}
0&1&w&w^{2}\\
1&0&w^{2}&w\\
w&w^{2}&0&1\\
w^{2}&w&1&0
\end{pmatrix}
\label{eq:gammaM}
\end{eqnarray}
where $w = \textrm{e}^{\imath 2\pi/3}$.

%======================================================
%======================================================
\section{Moessner-Chalker criterion}
\label{app:MCcriterion}

The Moessner-Chalker criterion provides a measure of frustration strength for a family of antiferromagnetic $O(n)$ models, using a Maxwell counting argument \cite{Moessner1998}. For a corner-sharing lattice, made of $N$ units (here triangles or tetrahedra) containing $q$ spins each, the number of ground-state degrees of freedom is $D_{M}=N[q(n-1)/2-n]$.

This criterion reproduces the extensive degeneracy of the Heisenberg pyrochlore ($q=4, n=3$) and brings the XY pyrochlore ($q=4, n=2$), Heisenberg hyperkagome ($q=3, n=3$) and Heisenberg kagome ($q=3, n=3$) to a marginal value, $D_{M}=0$ [\onlinecite{Moessner1998}]. This marginal value accounts for the fact that, despite an extensively degenerate ground-state, order-by-disorder induces a finite-temperature transition in the two former three-dimensional models \cite{Champion04a,Zhitomirsky2008}, while Mermin-Wagner-Hohenberg theorem pushes this transition to zero temperature in the two-dimensional Heisenberg kagome antiferromagnet \cite{Chalker1992,Zhitomirsky2008,Chern13b}.
\end{appendix}
%======================================================
%======================================================

\bibliographystyle{apsrev4-1}
\bibliography{library}
\end{document}